\documentclass[journal]{IEEEtran}
\usepackage{amsmath,amssymb,physics,tabularx,graphicx,multirow,hhline,algorithm,mathtools,algpseudocode,varwidth,diagbox}
\usepackage{lipsum}
\usepackage{color, soul}
\usepackage{xcolor}
\usepackage{empheq}
\usepackage[skins,breakable,hooks,theorems]{tcolorbox}
\usepackage{color, colortbl}

\tcbset{
  highlight math style={
    colback=yellow,
    arc=0pt,
    outer arc=0pt,
    boxrule=0pt,
    top=2pt,
    bottom=2pt,
    left=2pt,
    right=2pt,
  }
}

\ifCLASSINFOpdf
\else
\fi

\ifCLASSOPTIONcompsoc \usepackage[caption=false,font=normalsize,labelfont=sf,textfont=sf]{subfig}
\else
\usepackage[caption=false,font=footnotesize]{subfig}
\fi

\makeatletter
\def\ps@IEEEtitlepagestyle{%
  \def\@oddfoot{\mycopyrightnotice}%
  \def\@evenfoot{}%
}
\def\mycopyrightnotice{%
  {\hfill \footnotesize \copyright 2018 IEEE\hfill}
}
\makeatother

\begin{document}

\title{An Online Detection Framework for Cyber Attacks on Automatic Generation Control}

\author{Tong Huang,~\IEEEmembership{Student Member,~IEEE,}
		Bharadwaj Satchidanandan,
		P. R. Kumar,~\IEEEmembership{Fellow,~IEEE,}\\
        and Le Xie,~\IEEEmembership{Senior Member,~IEEE}
        \thanks{
        This material is based upon work partially
supported by NSF under Contract Nos. CNS-1646449, ECCS-1760554, and Science \& Technology Center Grant CCF-0939370, and the
U.S. Army Research Office under Contract No. W911NF-15-1-0279, and the Power Systems Engineering Research
Center (PSERC). \emph{(Corresponding author: Le Xie.)}

The authors are with the Department
of Electrical and Computer Engineering, Texas A\&M University, College Station,
TX, 77843 USA e-mail: tonghuang@tamu.edu, bharadwaj.s1990@tamu.edu, prk@tamu.edu, and le.xie@tamu.edu.}
        }



\maketitle

\begin{abstract}
We propose an online framework to detect cyber attacks on Automatic Generation Control (AGC). A cyber attack detection algorithm is designed based on the approach of Dynamic Watermarking. The detection algorithm provides a theoretical guarantee of detection of cyber attacks launched by sophisticated attackers possessing extensive knowledge of the physical and statistical models of targeted power systems. The proposed framework is practically implementable, as it needs no hardware update on generation units. The efficacy of the proposed framework is validated in both four-area system and 140-bus system.

\end{abstract}

\begin{IEEEkeywords}
Dynamic watermarking, cyber-physical security (CPS), automatic generation control (AGC).
\end{IEEEkeywords}

\IEEEpeerreviewmaketitle{}

\section{Introduction} 
\label{sec:introduction}
	The role of Automatic Generation Control (AGC) in large power systems is indispensable. It maintains nominal frequency while minimizing generation costs. The operation of the AGC involves close interaction between the cyber and the physical layers. By tracking Area Control Error (ACE) deviation collected from distributed sensors, the power outputs of generators are modified via AGC to balance random fluctuation of loads, and the electric grid frequency is thereby maintained within a tight range around the nominal value (50/60 Hz). However, due to the consequent tight coupling between the cyber and physical layers, there arises a vulnerability in that both grid stability and security can be compromised by malicious attacks on the cyber layer for sensing. Rather than compromising the strongly secured cyber layers of the control centers, cyber attacks on distributed measurements feeding the AGC might in fact significantly disrupt the operational goals of the power system \cite{7867825}.
	There have been several efforts at examining the potential mechanisms by which such cyber attacks on AGCs can be carried out and their negative impacts on the system operation. For example, as described in \cite{6740883}, several attempts for cyber attacks on AGCs, namely, scaling, ramp, pulse, and random attacks, may compromise both the physical system stability and the electricity market operation.
	Experiments based on CPS testbeds suggest that the corrupted measurements feeding the AGC might bring power systems to under-frequency condition and cause unnecessary load shedding \cite{7286615}, \cite{7573319}.
	By replacing the original measurements with an ``optimal attack sequence'', the malicious attackers can disrupt the system frequency in the shortest time without triggering certain pre-defined data quality alarms \cite{7867825}. Besides cyber attacks on AGC, potential risks can also be posed from the load side: adversaries may be able to trip targeted generators by manipulating the controller parameters of the loads offering emulated inertia control services \cite{656009}, \cite{7911324}. This paper focuses on the detection of cyber attacks on AGC.
	
	All of the above attack strategies on AGC are based on the assumption that the cyber layer of the AGC transporting the physical measurements is vulnerable to attacks, so that a malicious adversary can manipulate these measurements. Unfortunately, this assumption is validated by several recent real-world incidents. 
	Examples include computer viruses such as Dragonfly \cite{dragonfly} and Stuxnet \cite{Stuxnet} targeting Industrial Control Systems (ICS).
	Therefore, although no real-world attack specifically targeting the AGC has been reported thus far, the aforementioned attack strategies on AGC are more than theoretical concerns. As grid operation becomes more and more data-dependent, it is imperative to prepare the operators with an online defense mechanism against all possible cyber attacks on AGCs. 
	
	There have been several detection techniques for cyber attacks on AGCs. 
	In \cite{6740883}, cyber attacks following predefined attack strategies are detected by checking the statistical and temporal characterization of area control errors (ACE). 
	In \cite{6997497}, a statistical model learned from frequency and tie-line flow measurements is exploited to predict their short-term values. Measurements in the vicinity of their corresponding predictions are tagged as normal measurements. Otherwise, alarms are triggered. 
	In \cite{7867825}, the compromised tie-line flow measurements are detected by capturing the discrepancy between the meter readings of frequency deviation and its predicted value based on reported tie-line flow measurements and an identified linear-regression model.
	Also, DC state estimation (SE) is modified to be executed every AGC cycle and serves as an additional layer for data purification in \cite{7867825}.

	Although the aforementioned approaches increase the attack costs to some extent, 
	the measurements feeding the AGC may still be compromised by an attacker equipped with the following capabilities.
	First, the malicious adversaries are not constrained to follow the prescribed attack templates in order to cause significant impact on the grid \cite{7867825}. Although the anomaly detection engine proposed in \cite{6740883} is capable of identifying the predefined attack templates, there is no theoretical guarantee that the proposed algorithm can detect arbitrary cyber attacks. Second, extensive information on the system model might be exposed to the adversary. There are two ways by which a malicious adversary can obtain information about the power system model: 1) The detailed physical model may be directly leaked to the attacker via disgruntled employees or malicious insiders \cite{7936473}; 2) The statistical model of the power system can be learned using mathematical tools based on the leaked system operating data.
	 The attackers in the former case can bypass the SE-based detection algorithm by conducting ``unobservable attacks'' described in \cite{Liu:2009:FDI:1653662.1653666} or by conducting the packet-reordering integrity attack reported in \cite{7554505}, whereas the adversaries in the latter case can tamper with the measurements without triggering the alarm defined in \cite{6997497} by replacing the actual measurement sequence with a different sequence that still conforms to the learned statistical model \cite{7738534}. 
	Besides, the authors of \cite{7867825} exclude the attacks on frequency sensors from their framework.
	Therefore, a subtle but malicious distortion of frequency measurements based on the physical/statistical model of the power system is not likely to be detected by the algorithm proposed in \cite{7867825}.

	In this paper, we introduce a first-of-its-kind online detection framework of false data injection attacks in power systems. The recent dynamic watermarking technique \cite{7738534}, \cite{7945354} is employed in the framework and serves as the core algorithm to detect any tampered measurements feeding the AGC. Through deliberately superimposing a private signal of small magnitude upon the control commands sent by the AGC, we ``watermark'' the measurements feeding the AGC with certain indelible characteristics \cite{7738534}, by which cyber attacks on the AGC can be identified. To the best of the authors' knowledge, this is the first time that the dynamic watermarking technique has been applied to address cyber-security issues in power systems. The proposed framework has the following advantages.
	1) The detection algorithm used with the dynamic watermarking is theoretically rigorous and ensures that any manipulation of the measurements feeding the AGC can be detected regardless of the attack strategy that the attackers follow, as long as the controlled generators can execute commands from AGC honestly; 2) the algorithm can be used when attackers possess detailed information of the physical/statistical models of the power system; 3) the proposed framework is practically implementable, as it needs no hardware update on generation units.

	The rest of this paper is organized as follows. Section \ref{sec:problem_formulation} formulates the problem of detection of cyber attacks by  mathematically describing a system equipped with AGC and by presenting typical attack models; Section \ref{sec:dynamic_watermarking_based_defense_methodology} presents the dynamic watermarking-based detection algorithm in the context of AGC; Section \ref{sec:illustrative_example} validates the efficacy of the proposed algorithm via an illustrative example; Section \ref{sec:conclusion} concludes the paper.

\section{Problem Formulation} 
\label{sec:problem_formulation}
	In this section, a power system equipped with multiple AGCs is described mathematically, and typical attack templates 
	are presented.
	\subsection{The Model of a Multi-area Power System without AGC} 
	\label{sub:state_space_model}
		The dynamics of a multi-area power system in the vicinity of an operating condition can be described approximately by a continuous state-space model \cite{8116598}: 
		\begin{subequations} \label{eq: state-space-continuous}
			\begin{align}
				\boldsymbol{\dot{x}}(t)&=A\boldsymbol{x}(t)+B\boldsymbol{u}(t)+\boldsymbol{\gamma}'(t),\\
				\boldsymbol{y}(t)&=C\boldsymbol{x}(t)+\boldsymbol{n}'(t),
			\end{align}
		\end{subequations}
		where $\boldsymbol{x}(t)\in\mathbb{R}^{n'\times 1}$, $\boldsymbol{u}(t)\in\mathbb{R}^{d\times 1}$ and $\boldsymbol{y}(t)\in\mathbb{R}^{m\times 1}$ are states, inputs and measurements vectors in the time instant $t$, respectively, and the matrices $A$, $B$ and $C$ are system parameters of appropriate dimensions. Above $\boldsymbol{\gamma}'(t)\sim \mathcal{N}(0,Q')$ and $\boldsymbol{n}'(t)\sim \mathcal{N}(0,R')$ denote the white process noise and the measurement noise respectively that are independent of each other (A more mathematical description would entail stochastic differential equations). Suppose that there are $r$ control areas. Then, the measurement vector $\boldsymbol{y}(t)$ in \eqref{eq: state-space-continuous} can be reorganized as
			$\boldsymbol{y}(t) = 
						\begin{bmatrix}
							\boldsymbol{y}_{1}(t)^T &\boldsymbol{y}_{2}(t)^T & \cdots & \boldsymbol{y}_{i}(t)^T \cdots & \boldsymbol{y}_{r}(t)^T
						\end{bmatrix}^T$,
		where $(\cdot)^T$ is the transpose operation, and $\boldsymbol{y}_i(t)$ is a column vector incorporating all tie-line flow deviations $\boldsymbol{p}_{ti}(t),$ as well as the frequency deviation $\omega_i(t)$ in the control area $i$, i.e., 
		\begin{equation} \label{eq:y_i}
			\boldsymbol{y}_i(t)=
		\begin{bmatrix}
			\boldsymbol{p}_{ti}(t)^T & \omega_i(t)
		\end{bmatrix}^T.
		\end{equation} Similarly, the variables in $\boldsymbol{u}(t)$ can be grouped area-wise into
			$\boldsymbol{u}(t) = 
						\begin{bmatrix}
							\boldsymbol{u}_{1}(t)^T &\boldsymbol{u}_{2}(t)^T & \cdots & \boldsymbol{u}_{i}(t)^T \cdots & \boldsymbol{u}_{r}(t)^T
						\end{bmatrix}^T$,
		where the column vector $\boldsymbol{u}_i(t)$ includes the load reference setpoints $\boldsymbol{p}_{\text{s}i}(t)\in\mathbb{R}^{d'\times 1}$ of all generators participating in AGC in the area $i$, as well as local load fluctuation $\boldsymbol{p}_{\text{load}i}(t)+j\boldsymbol{q}_{\text{load}i}(t)$ at time instant $t$, i.e., 
		\begin{equation} \label{eq: p_ref}
			\boldsymbol{u}_i(t)=
		\begin{bmatrix}
			\boldsymbol{p}_{\text{s}i}(t)^T & \boldsymbol{u}_{\text{load}i}(t)^T
		\end{bmatrix}^T,	
		\end{equation} 
		where $\boldsymbol{u}_{\text{load}i}= 
		\begin{bmatrix}
			\boldsymbol{p}_{\text{load}i}(t)^T & \boldsymbol{q}_{\text{load}i}(t)^T
		\end{bmatrix}^T$.
	\subsection{The Model of a Multi-area System Regulated by AGC} 
	\label{sub:the_model_of_multi_area_system_regulated_by_agc}
		From a system-theoretic perspective, the AGC can be regarded as a multi-variable feedback loop added to the plant described in \eqref{eq: state-space-continuous}. In order to achieve independent regulation for the local tie-line flows and frequency, the Balancing Authority in one area only actuates the local generators participating in AGC without interference from generators in other areas. Therefore, the multi-area control policy can be decentralized area-wise as
		\begin{equation} \label{eq: control_policy}
			\begin{aligned}
				&\mathbf{u}[t]=\boldsymbol{f}(\boldsymbol{y}^t)\\
					&=\begin{bmatrix}
						\boldsymbol{f}_1(\boldsymbol{y}_1^t)^T & \boldsymbol{f}_2(\boldsymbol{y}_2^t)^T& \cdots & \boldsymbol{f}_i(\boldsymbol{y}_i^t)^T & \cdots & \boldsymbol{f}_r(\boldsymbol{y}_r^t)^T
					\end{bmatrix}^T,
			\end{aligned}
		\end{equation}
		where $\boldsymbol{y}_i^t$ is the telemetered measurement sequence up to time $t$ at area $i$. To elaborate on the control policy, suppose that there are $\psi$ local generation units in the AGC and $\phi$ measurements in area $i$, then the control policy of AGC $\boldsymbol{f_i}(\boldsymbol{\cdot}): \mathbb{R}^{\phi} \rightarrow \mathbb{R}^{\psi}$
		consists of the following operations between two successive economic dispatches: 
		\begin{enumerate}
			\item Area control error (ACE) is calculated from the telemetered tie-line flows and frequency measurements sampled every two to four seconds as
					$$ACE_i = \sum_{s=1}^{\phi}p_{ti,s} + \beta_i \omega_i,$$
			where the adjustable parameter $\beta_i$ is a bias factor.
			\item The ACE is smoothed by passing it through a low-order filter in order to mitigate the fatigue of generation control devices, e.g., turbine valves and governor motors \cite{kundur1994power}.
			\item At the balancing authority, a control command is computed from the ACE according to the control policy reported in \cite{1388528}, and is executed every two to four seconds \cite{kundur1994power}, \cite{7936473}. Denote by $\kappa_i\tau$ the time period between two consecutive commands.
			\item The control command computed by AGC is sent to the $\psi$ local generation units and its magnitude for each controlled generator is proportional to the coefficient updated by the economic dispatch algorithm \cite{carpentier1985or}, \cite{6025191}. 
		\end{enumerate}
The	above procedure (also summarized in Fig. \ref{fig:flowchart}) indicates that only the measurements at the chosen sample instants contribute to the computation of the control commands sent by the AGC at area $i$. The sequence $\boldsymbol{y}_i^t$ formed by these measurements is denoted by
		\begin{equation}
			\boldsymbol{y}_i^t:=\left\{\boldsymbol{y}_i(0), \boldsymbol{y}_i(\kappa_i\tau),\cdots, \boldsymbol{y}_i\left(\left \lfloor{\frac{t}{\kappa_i\tau}}\right \rfloor \kappa_i\tau \right)\right\}
		\end{equation}
		where $\left \lfloor{\cdot}\right \rfloor$ is the floor function.
		The above control policy yields the load reference setpoints $\boldsymbol{p}_{\text{s}i}(t),$ so that
		\begin{equation}\label{eq: close_loop}
			\boldsymbol{p}_{\text{s}i}(t) = \boldsymbol{f_i}(\boldsymbol{y}_i^t) \quad \forall i\in\{1, 2, \cdots, r\}.
		\end{equation}
		The above equation couples the physical infrastructure (generation units) and the cyber layer (control centers) together.
		In summary, \eqref{eq: state-space-continuous}, \eqref{eq: control_policy} and \eqref{eq: close_loop} constitute a hybrid model for a multi-area power system regulated by AGCs.

		Note that a commercial-level AGC includes more functions, which are assumed to be included into the control law \mbox{$\boldsymbol{f}_i(\cdot)$}. Fig. \mbox{\ref{fig:flowchart}} shows a simplified version of a realistic AGC.
		
	\subsection{Discretization of the Hybrid AGC Model} 
	Suppose that the time period between two consecutive control commands of AGC in each area is an integer multiple of a sampling time $\tau$, namely, $\kappa_i$ is assumed to be an integer. Then the continuous-time state space model \eqref{eq: state-space-continuous} can be discretized at $\tau$ using the approach reported in \cite{bay1999fundamentals}. For the sake of convenience, the discrete state-space model is denoted as $\mathcal{M}''_{\text{d}}$. Similarly, the AGC control policies in area can also be sampled at $\tau$. Denote the discrete control policies by $\boldsymbol{f}_{\text{d}i}(\cdot)$ for all $i\in\{1,2,\cdots,r\}$. It is worth noting that all areas are sampled with the same interval $\tau$, and the AGC in area $i$ sends control signals only after every $\kappa_i\tau$ seconds, for $i\in\{1,2,\cdots,r\}$. 

\begin{figure}[h!]
				\centering
				\includegraphics[width = 3.5in]{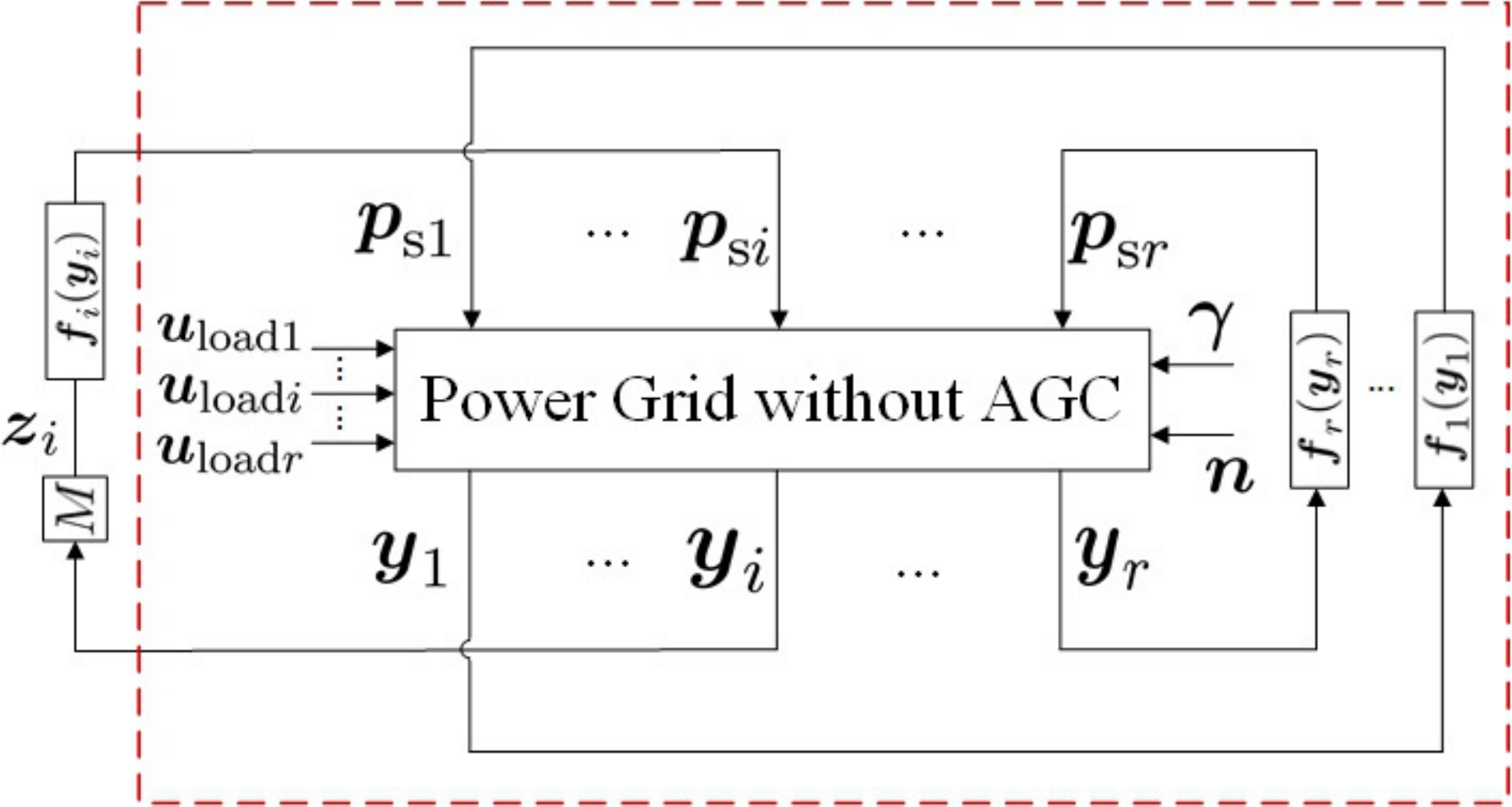}
				\caption{A multi-area power system with AGC systems.}
				\label{fig:feedback}
			\end{figure}

	For the control area $i$, we temporarily open its AGC feedback loop and keep the AGC loops in other areas $j$ connected, for $j\in\{1,2,\cdots,r\}$ and $j\ne i$. As shown in Fig. \ref{fig:feedback}, we focus on modeling the open-loop behavior of the system for area $i$ in terms of its inputs, i.e., the setpoints $\boldsymbol{p}_{\text{s}i}$ of the controlled generators in the area $i$, and all load fluctuations $\boldsymbol{u}_{\text{load}j}$ for all $j\in\{1,2,\cdots,r\}$, and its outputs, i.e., all tie-line flow deviations $\boldsymbol{p}_{ti}$ and frequency deviations $\omega_i$ in \eqref{eq:y_i}. As is standard in linear control theory \cite{feedback_matlab}, the discrete model of the aforementioned open-loop system can be obtained by interconnecting the entire system model $\mathcal{M}''_{\text{d}}$ and the discrete AGC control policies $\boldsymbol{f}_{\text{d}j}(\cdot)$, where $j\in\{1,2,\cdots,r\}$ and $j\ne i$. Denote the resulting interconnected state-space model for area $i$ by $\mathcal{M}'_{\text{d}i}$. It is worth noting that the state variables of $\mathcal{M}'_{\text{d}i}$ include all state variables in both state-space model $\mathcal{M}''_{\text{d}}$ and discrete control policies $\boldsymbol{f}_{\text{d}j}$, where $j\in\{1,2,\cdots,r\}$ and $j\ne i$. We specify setpoint $\boldsymbol{p}_{\text{s}i}$ as the control inputs of system $\mathcal{M}'_{\text{d}i}$, and further assume $\mathcal{M}'_{\text{d}i}$ is stabilizable \cite{bay1999fundamentals}. Finally, the discrete state-space model $\mathcal{M}'_{\text{d}i}$ can be minimally realized by a controllable and observable model $\mathcal{M}_{\text{d}i}$ with reduced order \cite{bay1999fundamentals}, namely,

	\label{sub:discretization_of_the_hybrid_agc_model}
		\begin{subequations} \label{eq: state-space-discrete_area_i_hybrid}
			\begin{align}
				&\begin{aligned}
					\boldsymbol{x}_{\text{d}i}(k+1)=&A_{\text{d}i}\boldsymbol{x}_{\text{d}i}(k)+B_{\text{d}i}^{\text{ref}}\boldsymbol{p}_{si}(k)\\
					&+ B_{\text{d}i}^{\text{load}}\boldsymbol{u}_{\text{load}}(k) + \boldsymbol{\gamma}(k+1)
				\end{aligned}
				\\
				&\boldsymbol{y}_i(k)=C_{\text{d}i}\boldsymbol{x}_{\text{d}i}(k) + \boldsymbol{n}(k)
			\end{align}
		\end{subequations}
	 where $\boldsymbol{x}_{\text{d}i}\in\mathbb{R}^{n\times 1}$ collects all state variables in the reduced-order model $\mathcal{M}_{\text{d}i}$ and $\boldsymbol{u}_{\text{load}}(k) = \begin{bmatrix}
			\boldsymbol{u}_{\text{load}1}^T & \boldsymbol{u}_{\text{load}2}^T &\cdots & \boldsymbol{u}_{\text{load}r}^T
		\end{bmatrix}^T$. Vector $\boldsymbol{\gamma}(t)\sim \mathcal{N}(0,Q)$ and $\boldsymbol{n}(t)\sim \mathcal{N}(0,R)$ are the white process and measurement noises, where $R$ is positive definite. 
		We assume that the rank of matrix $C_{\text{d}i}B_{\text{d}i}^{\text{ref}}$ equals $\phi$, which is the number of rows of $C_{\text{d}i}$.

	\subsection{Cyber Attack Models and Their Impacts} 
	\label{sub:attack_models}
		Due to the close interaction between the AGC and the generation units indicated by \eqref{eq: close_loop}, the adversary can compromise the physical layer of the power system by distorting the measurements $\boldsymbol{y}^t$. Denote by $\boldsymbol{z}^t$ the measurements \emph{reported} by the sensors. The sensors are supposed to report the actual value measured, i.e., they are supposed to report truthfully with $\boldsymbol{z}^t=\boldsymbol{y}^t$. However, an adversarial sensor might declare values that are different from the actual measurements, so that $\boldsymbol{z}^t\ne\boldsymbol{y}^t$. The purpose of this paper is to detect the inconsistency between the actual and the reported measurements caused  deliberately by the attacker. The attackers are assumed to be able to manipulate the distributed sensors feeding into AGC, i.e., frequency and tie-line flow measurements. Before describing the remedy for the problem, we present three typical attack templates.


		\subsubsection{Replay Attack} 
		\label{ssub:replay_attack}
			Before the attack, the adversary records the measurements during normal operating condition for some duration. During the attack, the actual measurements observed by the adversarial sensors are replaced by the recorded measurements and reported to the control center \cite{5394956}. 

		\subsubsection{Noise-injection Attack} 
		\label{ssub:noise_injection_attack}
			Under this attack model, the adversarial sensors add a bounded random value to the actual measurement and then report it to the control center.
		\subsubsection{Destabilization Attack} 
		\label{ssub:destabilization_attack}
			In a destabilization attack, the compromised sensors of the AGC in area $i$ report a sequence $\{\boldsymbol{z}_i\}$ which is a filtered version of the actual measurement sequence $\{\boldsymbol{y}_i\}$. If $M$ denotes such a filter, the attack consists of inserting the filter $M$ to the system model, with $M$ so chosen such that the original system becomes unstable. It is worth noting that the output sequence \mbox{$\boldsymbol{z}_i$} of a malicious filter \mbox{$M$} can be obtained through a simple tuning procedure, even without any information on the system model, as will be described in Section \mbox{\ref{subs:detection_of_destabilization_attack}}.

			Note that the attackers are not limited to follow any attack templates, in their attempt to bring harm to power systems. Correspondingly, a defense method should be designed not only for detecting the three types of attacks defined above, but also for securing AGC from any manipulation on the distributed measurements feeding into AGC.
\section{Dynamic Watermarking-based Defense} 
\label{sec:dynamic_watermarking_based_defense_methodology}
	In this section, we apply the approach of dynamic watermarking reported in \cite{7738534}, \cite{7945354}, \cite{6612700} to secure the distributed measurements feeding AGC in power systems. The fundamental idea of Dynamic Watermarking is as follows. The actuators (generation units in this case) superimpose on the control policy-specified input, a ``small" random signal chosen according to a certain probability distribution. While this probability distribution is made public, so that even the adversary knows it, the actual realization of the random signal is known only to that particular generation unit, and it doesn't reveal that to any other party. For this reason, the random signal is also called the private excitation of the generators. In such a scenario, the honest sensors and the malicious sensors are distinguished by the following fact: the truthful measurements reported by the honest sensors exhibit certain expected statistical properties that are relevant to the statistics of the private excitation, whereas, as shown in \cite{7738534}, \cite{7945354}, measurements reported by the malicious sensors, if excessively distorted, do not exhibit these properties. Therefore, by subjecting the reported measurements to certain tests for these statistical properties, malicious activity in the system can be detected. 

	In this paper, we will demonstrate the application of this approach in the context of power systems. For control area $i$, an independent and identically distributed (i.i.d.) private excitation $\{\boldsymbol{e}_i(k)\}$ is superimposed on the control inputs $\{\boldsymbol{p}_{si}(k)\}$ \cite{7945354}. Consequently, the input applied at time $k$ is 
	\begin{equation} \label{eq: superposition}
		\boldsymbol{p}_{si}(k) = \boldsymbol{f}_i(\boldsymbol{y}_i^k)+\boldsymbol{e}_i(k),
	\end{equation}
	where $\boldsymbol{e}_i(k) \sim \mathcal{N}(0,\sigma_e^2I)$. It is worth noting that \eqref{eq: superposition} can be implemented by modifying the AGC software at the balancing authorities without any hardware updates on the generation units. With the private injection $\{\boldsymbol{e}_i(k)\}$, any attempt to distort the measurements fed to AGC will be detected by subjecting the reported measurements to the two tests \cite{7945354} described below. A detailed proof for this conclusion can be found in \mbox{\cite{7945354}}.
	\begin{figure}[h!]
			\centering
			\includegraphics[width=3.5in]{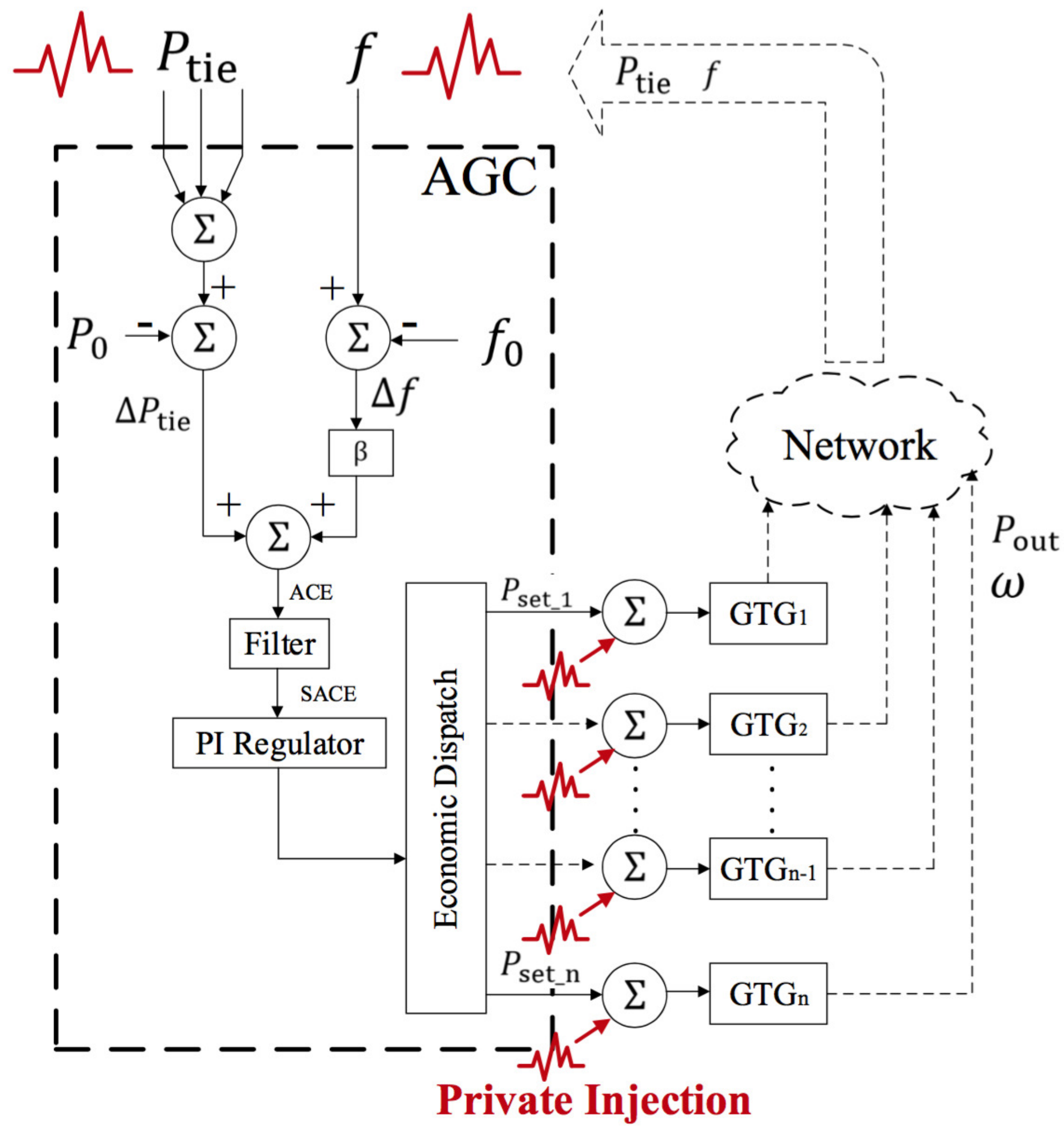}
			\caption{Location of Private Injection in a Simplified Functional Diagram of AGC}
			\label{fig:flowchart}
		\end{figure}
	\subsection{Two Indicators of Dynamic Watermarking} 
	\label{sub:dynamic_watermarking_technique}
	Given the input sequence $\boldsymbol{u}_i$ and measurement sequence $\boldsymbol{y}_i$ of the discrete system \eqref{eq: state-space-discrete_area_i_hybrid} up to the $k$th unit of time, the system state $\boldsymbol{x}_{\text{d}i}(k|k)$ can be estimated by the Kalman filter as
	\begin{subequations} \label{eq: Kalman}
	\begin{align}
		&\begin{aligned}
			\boldsymbol{x}_{\text{d}i}(k+1|k) &= A_{\text{d}i}(I-L_{\text{d}i}C_{\text{d}i})\boldsymbol{x}_{\text{d}i}(k|k-1)+\\&\begin{bmatrix}
		B_{\text{d}i}^{\text{ref}}&
		B_{\text{d}i}^{\text{load}}&
		A_{\text{d}i}L_{\text{d}i}
	\end{bmatrix}
	\begin{bmatrix}
		\boldsymbol{p}_{\text{s}i}(k)\\
		\boldsymbol{u}_{\text{load}}(k)\\
		\boldsymbol{y}_i(k)
	\end{bmatrix},
		\end{aligned}
		\\
	&\boldsymbol{x}_{\text{d}i}(k|k) = (I-L_{\text{d}i}C_{\text{d}i})\boldsymbol{x}_{\text{d}i}(k|k-1) + L_{\text{d}i}\boldsymbol{y}_i(k),
	\end{align}
\end{subequations}
where $L_{\text{d}i}$ is the steady-state Kalman filtering gain given by
	\begin{equation}
		L_{\text{d}i} = PC_{\text{d}i}^T(C_{\text{d}i}PC_{\text{d}i}^T+R)^{-1}.
	\end{equation}
	In the above, $P$ is obtained as the unique positive definite solution of the Algebraic Riccati Equation \cite{kumar2015stochastic}. 
	
	We define 
	\begin{equation} \label{eq: nu_k}
		\begin{aligned}
			\boldsymbol{\zeta}_k :=& \boldsymbol{x}_{\text{d}i}(k|k)-A_{\text{d}i}\boldsymbol{x}_{\text{d}i}(k-1|k-1)-B_{\text{d}i}^{\text{ref}}\boldsymbol{f}_i(\boldsymbol{z}_i^{k-1})\\&-B_{\text{d}i}^{\text{ref}}\boldsymbol{e}(k-1)
			-B_{\text{d}i}^{\text{load}}\boldsymbol{u}_{\text{load}}.
		\end{aligned}
	\end{equation}

		\subsubsection{Test 1} 
		\label{ssub:test_1}
			Check if
			\begin{equation} \label{eq: test_1_eq}
					\lim_{T\rightarrow\infty} \frac{1}{T}\sum^{T}_{k=1}\boldsymbol{\zeta}_k\boldsymbol{\zeta}_k^T=L_{\text{d}i}\Sigma_iL_{\text{d}i}^T,
			\end{equation}
		where 
		\begin{equation}
			\Sigma_i := C_{\text{d}i}PC_{\text{d}i}^T+R.
		\end{equation}
		Correspondingly, we choose a time window $T$ and define an indicator matrix $W$ by
		\begin{equation} \label{eq: xi1}
			W (T) := \frac{1}{T}\sum^{T}_{k=1}\boldsymbol{\zeta}_k\boldsymbol{\zeta}_k^T-L_{\text{d}i}\Sigma_iL_{\text{d}i}^T.
		\end{equation}

	\subsubsection{Test 2} 
	\label{ssub:test_2}
	 	Check if
	 	\begin{equation} \label{eq: test_2_eq}
	 		\lim_{T\rightarrow\infty} \frac{1}{T}\sum^{T}_{k=1}\boldsymbol{e}(k-1)\boldsymbol{\zeta}_k^T=0.
	 	\end{equation}
	 	As before, we define
	 	\begin{equation} \label{eq: xi2}
	 		V (T): = \frac{1}{T}\sum^{T}_{k=1}\boldsymbol{e}(k-1)\boldsymbol{\zeta}_k^T.
	 	\end{equation}
	This measure can be calculated by the system operators. The reported measurements of interest, $\{\boldsymbol{z}_i(k)\}$, will pass both tests if $\boldsymbol{z}_i(k)\equiv\boldsymbol{y}_i(k)$ for all $k$; if the sensors distort the measurements beyond adding a zero-power signal, then, as shown in \cite{7945354}, at least one of the above tests will fail. While tests \eqref{eq: test_1_eq} and \eqref{eq: test_2_eq} are asymptotic in nature, they can be converted to statistical tests that can be performed in finite time. 
	For example, we expect much bigger entries in $W$ or $V$ during cyber attacks, than their counterparts when no attack happens. This leads naturally to a threshold test for detecting malicious distortion.

	\subsection{Online Algorithm for Detection of Cyber Attacks} 
	\label{sub:proposed_algorithm}
	The computation of the aforementioned indicators requires a sequence of reported measurements $\{\boldsymbol{z}_i\}$, private injections $\{\boldsymbol{e}_i\}$, load fluctuations of the whole grid $\{\boldsymbol{u}_{\text{load}}\}$ and AGC command signals $\{\boldsymbol{f}_i(\boldsymbol{z}_i^{k-1})\}$ over a period of time. Therefore, in order to check whether the reported measurements pass the two tests \eqref{eq: test_1_eq}, \eqref{eq: test_2_eq}, the generation unit processes  a block of $\{\boldsymbol{z}_i\}$, $\{\boldsymbol{e}_i\}$, $\{\boldsymbol{u}_{\text{load}}\}$ and $\{\boldsymbol{f}_i(\boldsymbol{z}_i^{k-1})\}$ within a time window $T$. Suppose that each block of the above sequences includes $T$ samples. Then, up to time $t=j\times T\times \kappa_i\tau$, we will have $j$ blocks of above sequences. The $j$th block of above sequences in area $i$ are denoted by $\boldsymbol{z}_{i}^{\text{BL}j}$, $\boldsymbol{e}_i^j$, $\boldsymbol{u}_{\text{load}}^j$ and $\boldsymbol{f}_i^j$, respectively: 
	$$\boldsymbol{z}_{i}^{\text{BL}j}:= \{\boldsymbol{z}_i((j-1)T), \boldsymbol{z}_i((j-1)T+1),\cdots, \boldsymbol{z}_i(jT)\},$$ $$\boldsymbol{e}_i^j:= \{\boldsymbol{e}_i((j-1)T), \boldsymbol{e}_i((j-1)T+1),\cdots, \boldsymbol{e}_i(jT)\},$$ $$\boldsymbol{u}_{\text{load}}^j:=\{\boldsymbol{u}_{\text{load}}((j-1)T), \boldsymbol{u}_{\text{load}}((j-1)T+1), \cdots, \boldsymbol{u}_{\text{load}}(jT)\},$$ and $$\boldsymbol{f}_i^j:= \{\boldsymbol{f}_i(\boldsymbol{z}_i^{(j-1)T-1}), \boldsymbol{f}_i(\boldsymbol{z}_i^{(j-1)T}),\cdots, \boldsymbol{f}_i(\boldsymbol{z}_i^{jT-1})\}.$$ 
	In terms of online application, 
	let $W^j=[w_{g,h}^j]$ and $V^j=[v_{g,h}^j]$ be $W$ and $V$ calculated within the $jth$ time window, respectively.
	Then the indicator scalars $\xi_1^j$ and $\xi_2^j$ are defined as follows
	\begin{subequations}
		\begin{align}
			\xi_1^j &:= \abs{\trace(W^j)}\\
			\xi_2^j &:= \sqrt{\sum_{g=1}^{d'}\sum_{h=1}^{n}(v_{g,h}^j)^2}
		\end{align}
		\label{eq: two_indicators}
	\end{subequations}
	where $\trace(\cdot)$ is the trace operator. As mentioned in \eqref{eq:y_i} and \eqref{eq: state-space-discrete_area_i_hybrid}, $d'$ is the number of the controlled generators in AGC of area $i$ and $n$ is the order of the reduced-order model in \eqref{eq: state-space-discrete_area_i_hybrid}. Finally, we expect $\xi_1^j\ge \eta_1$ or $\xi_2^j\ge\eta_2$, if attacks are launched in the $jth$ time window, where $\eta_1$ and $\eta_2$ are pre-defined thresholds.
	The thresholds \mbox{$\eta_1$} and \mbox{$\eta_2$} can be obtained from the following training procedure:
	\begin{enumerate}
		\item based on \mbox{\eqref{eq: xi1}} and \mbox{\eqref{eq: xi2}}, first compute \mbox{$W^{\infty}=W(T_{\infty})$} and \mbox{$V^{\infty}=V(T_{\infty})$} under normal operating condition, where \mbox{$T_{\infty}$} is a large integer that is set to $1800$ in this paper;
		\item obtain the general indicators \mbox{$\xi_1^{\infty}$} and \mbox{$\xi_2^{\infty}$} under a normal condition by \mbox{\eqref{eq: two_indicators}};
		\item the thresholds \mbox{$\eta_1$} and \mbox{$\eta_2$} are calculated by
		\begin{equation}
			\eta_1 = \kappa'\xi_1^{\infty} \quad \eta_2 = \kappa'\xi_2^{\infty}
		\end{equation}
		where \mbox{$\kappa'$} is an empirically adjustable parameter.
	\end{enumerate}
	The detection thresholds \mbox{$\eta_1$}, \mbox{$\eta_2$} can also be determined using the Neyman-Pearson criterion based on the maximum tolerable false alarm rate, which is shown in Appendix \mbox{\ref{sec:determining_the_threshold_using_the_neyman_pearson_criterion}}.
	Algorithm 2 specifies the subroutine for computing the two indicators $\xi_1^j$ and $\xi_2^j$ for the $jth$ block of measurements.

	For area $i$, private signals $\boldsymbol{e}_i$ are superimposed upon the AGC commands according to \eqref{eq: superposition} and Fig. \ref{fig:flowchart}. Then Algorithm \ref{alg:1} enables the balancing authority of area $i$ to detect cyber attacks on the measurements feeding the AGC. Once attacks in area $i$ are detected, the balancing authority stops sending commands to the generators in the AGC. Similarly, attacks to other areas can be detected by the corresponding balancing authorities similarly equipped with Algorithm \ref{alg:1}. Additionally, it is worth emphasizing that Fig. \mbox{\ref{fig:flowchart}} is a \mbox{\emph{simplified}} functional diagram of AGC, where the optimal power setpoints are the actual outputs of the simplified AGC. In the proposed method, the private excitations \mbox{$\boldsymbol{f}_i(\boldsymbol{y}_i^k)$} are supposed to be superimposed upon the actual outputs  of AGC, which is not necessary to be the calculated optimal power setpoint in a realistic AGC.
	
		\begin{algorithm}
			\caption{Online Algorithm for Detection of Cyber Attack} \label{alg:1}
			\begin{algorithmic}[1]
				\State $H \gets L_{\text{d}i}\Sigma_iL_{\text{d}i}^T$; $j\gets1$
				\While{$k = 1,2,\cdots,$}
					\If{$k\ge jT$}
						\State Obtain the sequence $\boldsymbol{z}_{i}^{\text{BL}j}$, $\boldsymbol{e}_i^j$, $\boldsymbol{u}^j_{\text{load}}$, $\boldsymbol{f}_i^j$;
						\State 
						\begin{varwidth}[t]{\linewidth}
							Compute $\boldsymbol{x}_{\text{e}}:= \{\boldsymbol{x}(k'|k')\}$ by \eqref{eq: close_loop} and \eqref{eq: Kalman} for all \\$k'= (j-1)T, (j-1)T+1, \cdots, jT$;
						\end{varwidth}
						\State {$\xi_1^j, \xi_2^j \gets$ \texttt{Indicators}($\boldsymbol{x}_{\text{e}}^j$, $\boldsymbol{e}_i^j$, $\boldsymbol{u}^j_{\text{load}}$, $\boldsymbol{f}_i^j$, $j$, $H$)}; 
						\State $j\gets j+1$
						\If{$\xi_1\ge \eta_1 \lor \xi_2\ge \eta_2$}
						\State
							\begin{varwidth}[t]{\linewidth}
								Claim attacks and stop sending commands to \\the generators on AGC;
							\end{varwidth}
						\EndIf
					\EndIf
				\EndWhile
			\end{algorithmic}
		\end{algorithm}

		\begin{algorithm} 
			\caption{Computation of $\xi_1^j$ and $\xi_2^j$ at the $j$th block} \label{alg: alg_2}
			\begin{algorithmic}[1]
				\Function{\tt Indicators}{$\boldsymbol{x}_{\text{e}}^j$, $\boldsymbol{e}_i^j$, $\boldsymbol{u}^j_{\text{load}}$, $\boldsymbol{f}_i^j$, $j$, $H$}
				\State $\Sigma_{s1} \gets 0$; $\Sigma_{s2} \gets 0$
				\While{$k = (j-1)T, (j-1)T+1, \cdots, jT,$}
		        \State Compute $\boldsymbol{\zeta}_k$ by \eqref{eq: nu_k}
		        \State $\Sigma_{s1} \gets \Sigma_{s1}+ \boldsymbol{\zeta}_k\boldsymbol{\zeta}_k^T$; $\Sigma_{s2} \gets \Sigma_{s2}+\boldsymbol{e}(k-1)\boldsymbol{\zeta}_k^T$
		      	\EndWhile
		      	\State $W^j = \frac{1}{T}\Sigma_{s1}-H$; $V_2^j = \frac{1}{T}\Sigma_{s2}$
		      	\State Obtain $\xi_1^j$ and $\xi_2^j$ via \eqref{eq: two_indicators}
		      	\State \textbf{return} $\xi_1^j$, $\xi_2^j$
				\EndFunction
			\end{algorithmic}
		\end{algorithm}	

		After a cyber attack is detected by the proposed framework, the AGC should be deactivated. It is worth noting that an efficient procedure for finding malicious sensors should be initiated after deactivating the AGC. Such a procedure may include dispatching a panel to investigate the distributed measurements after an alarm. Also, the procedure is required to correct the malicious sensors quickly. This requirement is achievable due to the limited number of the distributed measurements feeding to AGC. After clearing the cyberattacks, the AGC should be back to service. Therefore, the AGC is actually absent only for a short period of time, instead of permanently out of service.

One might wonder if the temporary absence of AGC significantly impacts the system frequency. The answer is that the temporary absence of AGC should not be a big concern, as the AGC is allowed to be deactivated in real-world system operation during some situations such as intentional tripping of load/generation \mbox{\cite{kundur1994power}}. Even without fine adjustments of frequency owing to AGC, the primary frequency control is capable of maintaining the system frequency within an acceptable range, say, from 59.96 Hz to 60.04 Hz \mbox{\cite{EPRI1234}}, and the frequency falling into such a range will not trigger any load shedding events \mbox{\cite{NERC_LS}}. However, if stealthy cyberattacks on AGC are not detected in a timely fashion, they may keep compromising the control performance of the frequency regulation. For example, if a replay attack is not detected in time, the energy consumed by AGC actually bring no benefit to the grid in terms of regulating frequency, and the control performance of AGC is compromised.

\section{Numerical Examples} 
\label{sec:illustrative_example}
This section presents the results on the efficacy of the dynamic-watermarking-based online defense algorithm on a four-area power system and the Northeastern Power Coordinating Council (NPCC) 140-bus power system. The malicious attacks to the synthetic system will be launched based on the attack templates presented in Sec. \ref{sub:attack_models}. As will be shown, these attacks can be detected in a timely manner via the proposed approach without sacrificing the performance of the system.
	\subsection{Performance Validation of the Proposed Algorithm on the Four-area System} 
	\label{sub:result_of_the_framework_for_the_four_area}
	
	\subsubsection{Four-area System Description} 
	\label{subs:system_description}
\begin{figure}[h!]
		\centering
		\includegraphics[width = 3.5in]{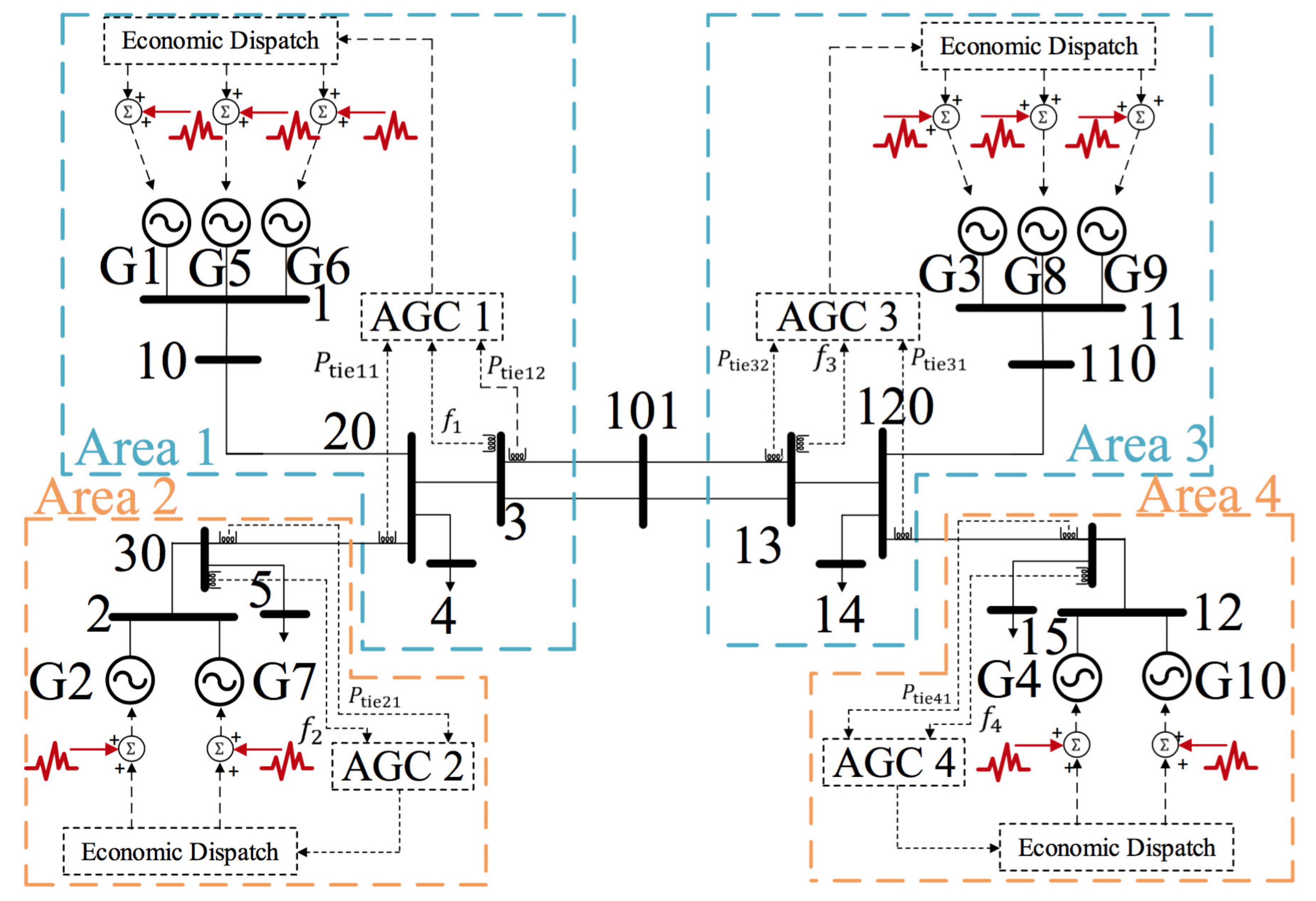}
		\caption{Four-area synthetic system with AGC in each area.}
		\label{fig:Kundur_system}
	\end{figure}

	This test system has four areas and ten generators, as shown in Fig. \ref{fig:Kundur_system}. The system is linearized about the given operating condition by Power System Toolbox (PST) \cite{chow1992toolbox}, and the system matrices for the linear model, i.e., $A$, $B$ and $C$ in \eqref{eq: state-space-continuous}, are extracted.
	In order to mimic the behavior of AGC, in each area, we add a discrete proportional-integral (PI) feedback loop, where the proportional gain constant is set to $-0.0745$ and the integral gain is set to $-0.0333$. For each area, the PI controller takes its local measurements of tie-line power flows and frequency as its inputs and computes a control signal to change the load reference setpoint of the generator. This is done every $2$ seconds, i.e., $\tau=2$ and $\kappa_i\equiv1$ for $i\in\{1,2,3,4\}$. 
	The load deviations 
	around the scheduled values are modeled as independent and identically distributed (i.i.d.) Gaussian white noise with zero mean and covariance matrix $\sigma_L^2 I_8$, where $I_8$ is a $8\times8$ identity matrix. The variance $\sigma_L^2=0.0025$ is chosen such that the frequency fluctuates within the normal range, i.e., $60 \pm 0.03$ Hz \cite{EPRI1234} with high probability. 
	The measurement noise of frequency and real power are normally distributed with zero mean. The variance of the frequency measurement noise, $\sigma_f^2=9.1891\times10^{-12}$, is tuned such that the accuracy of frequency measurement falls within $\pm 0.0005$ Hz \cite{5504177} with high probability, and the signal-to-noise ratio (SNR) of deviation measurements of tie-line flow is $20$ dB. The covariance matrix of the process noise $Q'$ is $10^{-9}I_{n'}$, 
		where $I_{n'}$ is an identity matrix of dimension of $n'$.

	\subsubsection{Parameter Setting of the Proposed Algorithm} 
	\label{subs:parameter_setting_of_the_proposed_algorithm}
		For the implementation of Algorithms \ref{alg:1} and \ref{alg: alg_2}, we have the following settings of the parameters:
		\begin{itemize}
			\item The number of samples in each block $T$ is 30, so that $\xi_1^j$ and $\xi_2^j$ are computed every $60$ seconds;
			\item The threshold \mbox{$\eta_1$} is set to \mbox{$2.5207\times 10^{-4}$} with \mbox{$\xi^{\infty}_1=3.6010\times10^{-5}$} and \mbox{$\kappa' = 7$};
			\item the variance of the private injections $\sigma_e$ in both Area $1$ and Area $2$ is set to $10^{-7}$.
		\end{itemize}

\begin{figure}[h!]
			\centering
			\includegraphics[width = 3.5in]{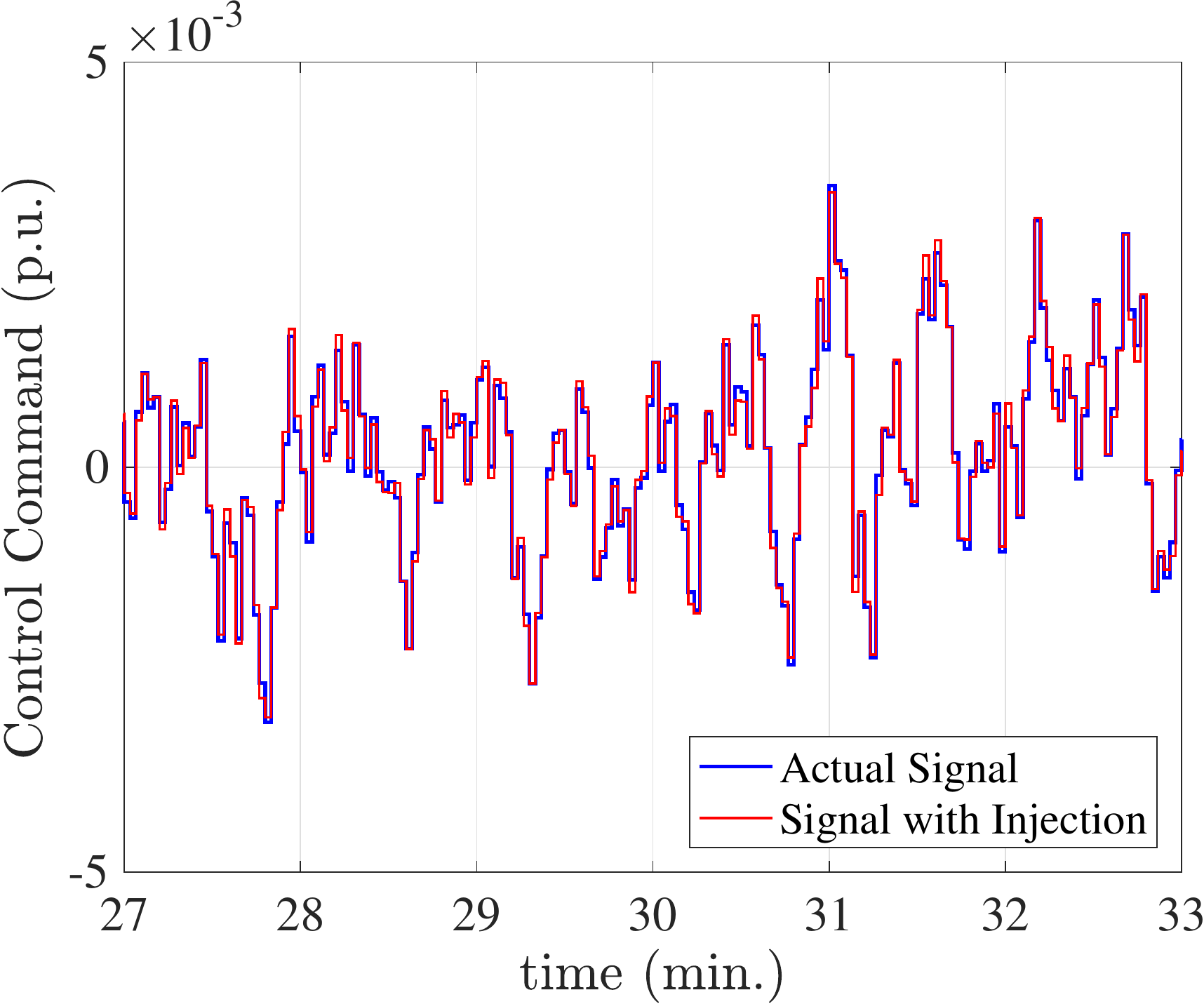}
			\caption{The impact of the private injection on the command signal showing that watermarking does not lead to any loss of performance under normal operation.}
			\label{fig:command_signal_record}
		\end{figure}

		We first examine the impact of the private injection on the performance of the AGC in terms of frequency regulation. Fig. \ref{fig:command_signal_record} records the control commands from AGC $1$, and it shows that the private injection does not cause significant deviation of the actual input from the control policy-specified input. 
		The percentage of variance change of control command of AGC 1 and frequency are $0.26\%$ and $1.73\%$, respectively, and the small change of the variance suggests negligible sacrifice of performance resulting from the private injection.

		%
		%
		%

	\subsubsection{Detection of Replay Attack} 
	\label{subs:detection_of_replay_attack}
		\begin{figure}[h!]
				\centering
				\subfloat[]{\includegraphics[width=1.7in]{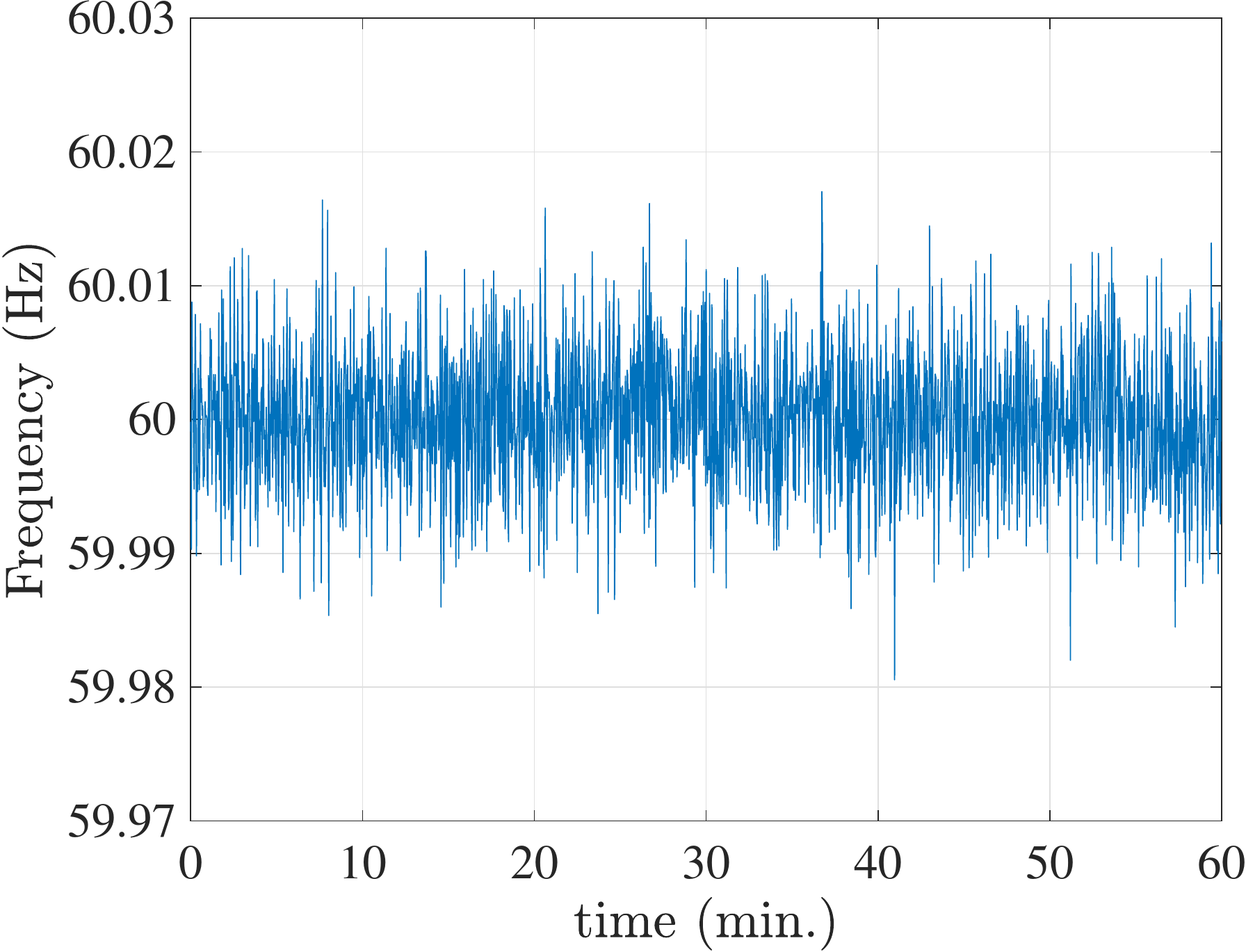}} 
				\hfil
				\subfloat[]{\includegraphics[width=1.7in]{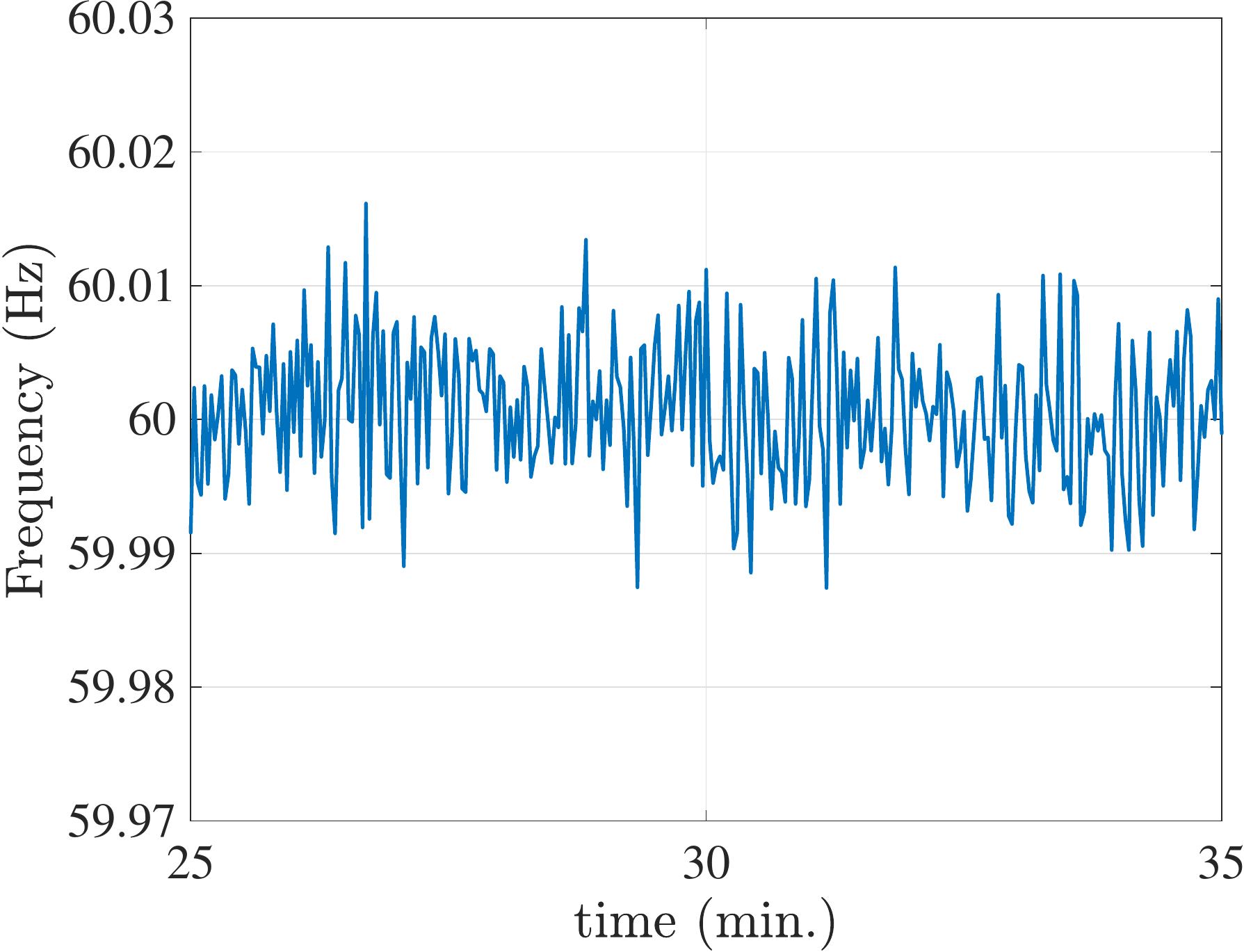}}
				\hfil
				\caption{Frequency measurement (a) from $0$ min to $60$ min, and (b) zoom-in frequency measurement from $25$ min to $35$ min, under the \textbf{replay} attack to the frequency measurement of Area 1 launched at $30$ min.}
				\label{fig: replay_measurement_freq_time_domain}
			\end{figure}
		We next demonstrate the efficacy of the dynamic watermarking approach for detecting replay attacks defined in Sec. \ref{ssub:replay_attack}. Figure \ref{fig: replay_measurement_freq_time_domain} shows the frequency measurements in Area $1$. Beginning at $30$ min, the frequency sensor reports a pre-recorded sequence of measurements instead of the actual measurements. No anomaly can be identified from Fig. \ref{fig: replay_measurement_freq_time_domain}, as no frequency constraint is violated within the time period of interest. 

\begin{figure}[h]
				\centering
				\subfloat[]{\includegraphics[width=1.7in]{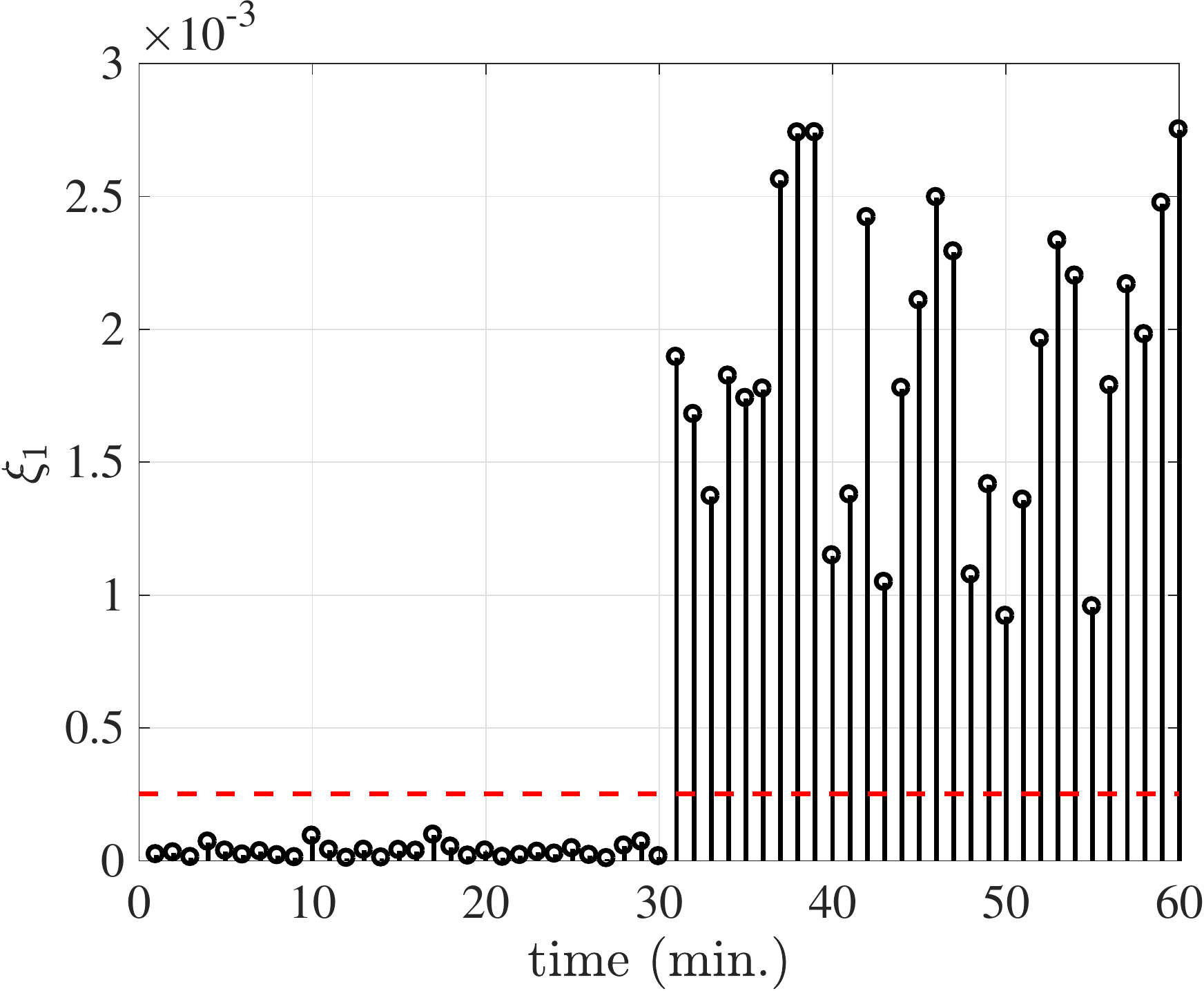}} 
				\hfil
				\subfloat[]{\includegraphics[width=1.7in]{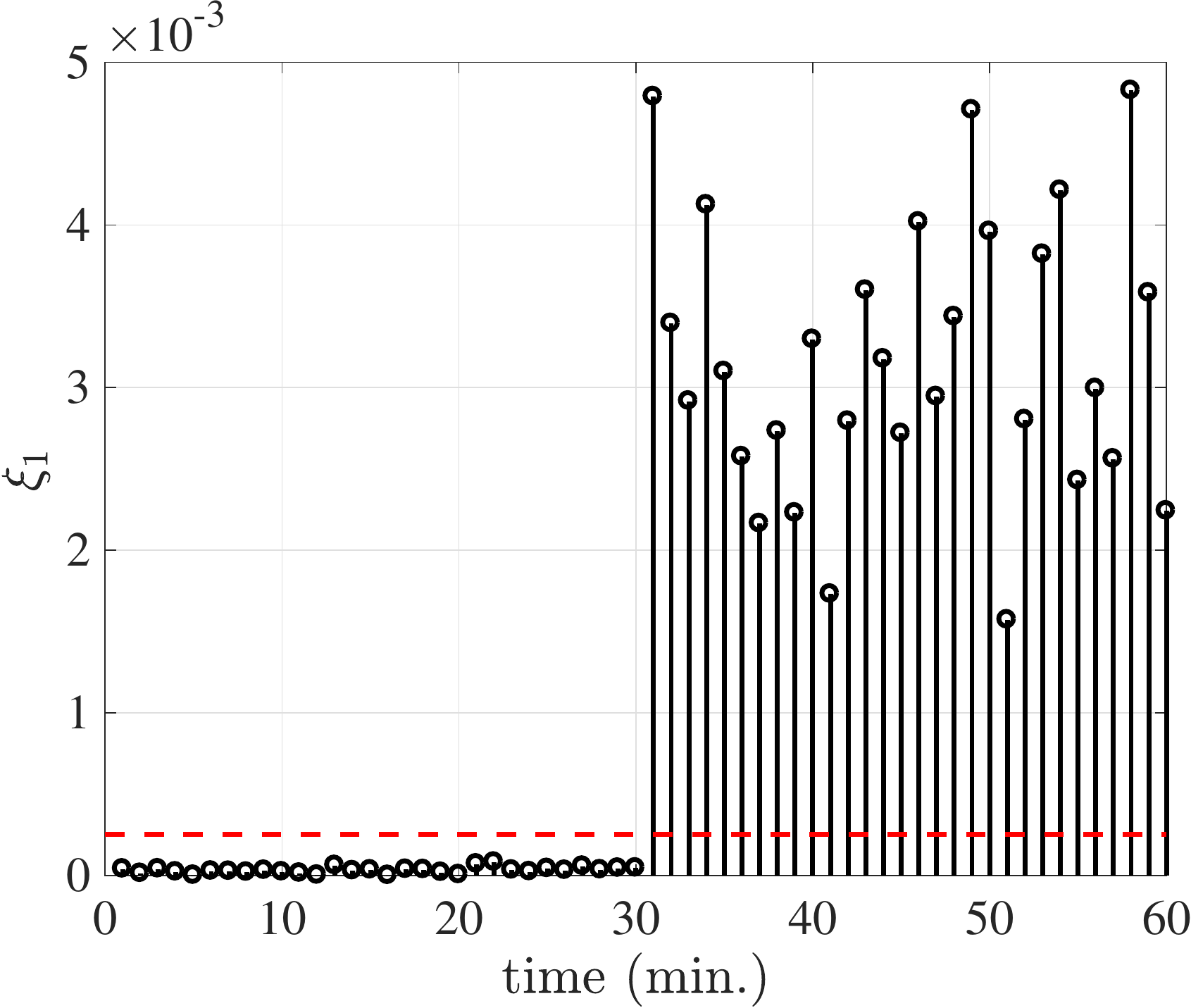}} 
				\hfil
				\caption{The evolutions of indicator $\xi_1^{\text{j}}$ under the \textbf{replay} attack to the (a) \textbf{frequency} measurement and (b) \textbf{tie flow} measurement of \textbf{Area 1} starting at $30$ min.}
				\label{fig: replay_detection_freq_area_1}
			\end{figure}

		Next, the proposed Algorithms 1 and 2 are applied to detect the replay attack. In each area, the online detection algorithms compute the indicators $\xi_1^j$ and $\xi_2^j$ based on their local measurements of frequency and tie-line flow. 
		The evolution of $\xi_1^j$ over time in Area $1$ is presented in Figure \ref{fig: replay_detection_freq_area_1}(a). It is seen that $\xi_1^j$ exceeds the threshold $\eta_1$ after $31$ minutes, indicating that the attack starts between the $30th$ and $31st$ minutes.
		A similar result can be observed from Fig. \ref{fig: replay_detection_freq_area_1}(b) which presents the evolution of $\xi_1^j$ under the replay attack to tie-line flow measurement of Area $1$. After the attacks are detected, one mitigation action is to deactivate the AGC.

	\subsubsection{Detection of Noise-injection Attack} 
	\label{subs:detection_of_noise_injection_attack}
\begin{figure}[h] 
				\centering
				\subfloat[]{\includegraphics[width=1.7in]{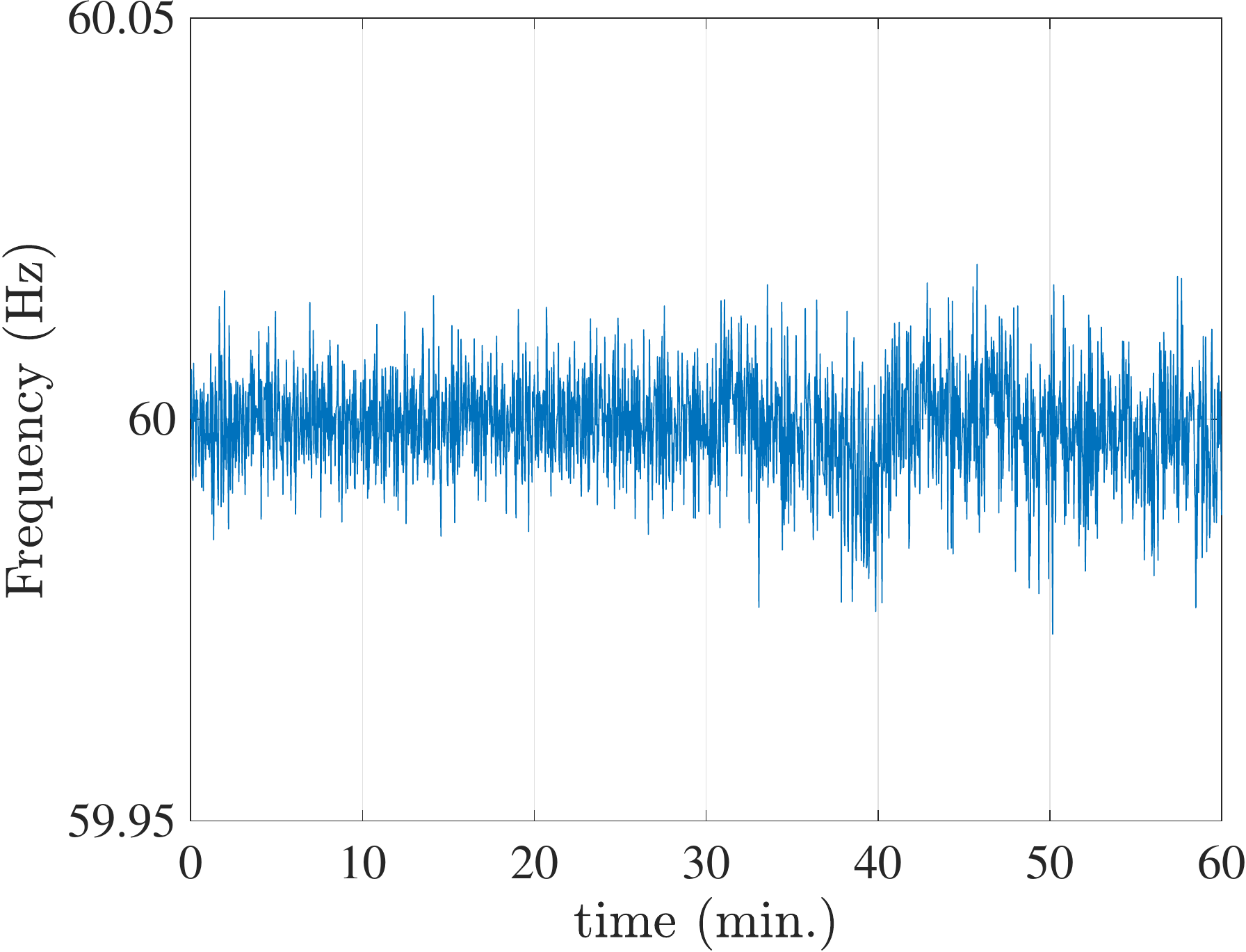}}
				\hfil
				\subfloat[]{\includegraphics[width=1.7in]{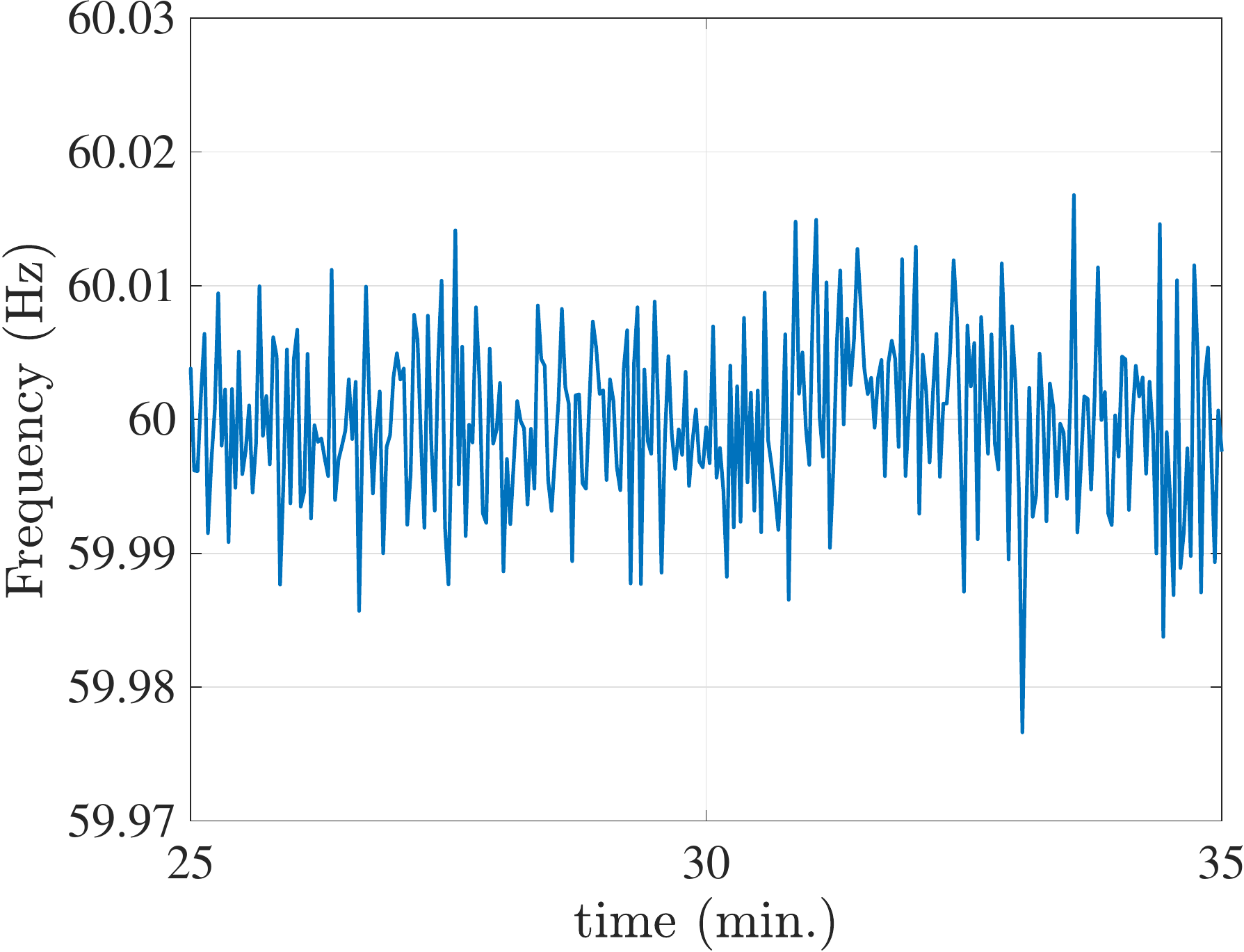}}
				\hfil
				\caption{Frequency measurement in Area $1$ (a) from $0$ min to $60$ min and (b) zoom-in frequency measurement from $25$ min to $35$ min, under the \textbf{noise-injection} attack to the frequency measurement of Area 1 launched at $30$ min.}
				\label{fig:noise_injection}
			\end{figure}

\begin{figure}[h!]
				\centering
				\subfloat[]{\includegraphics[width=1.7in]{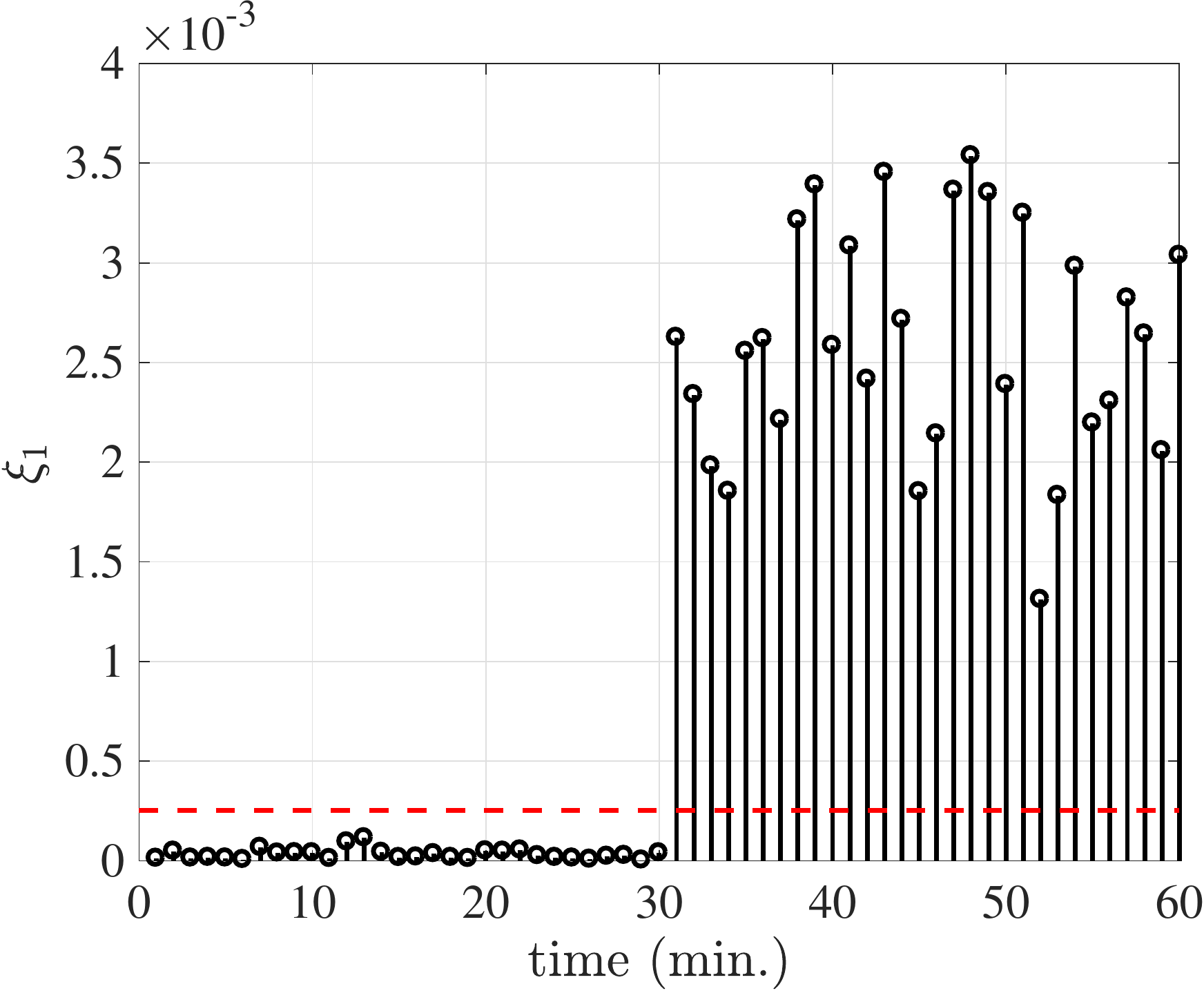}}
				\hfil
				\subfloat[]{\includegraphics[width=1.7in]{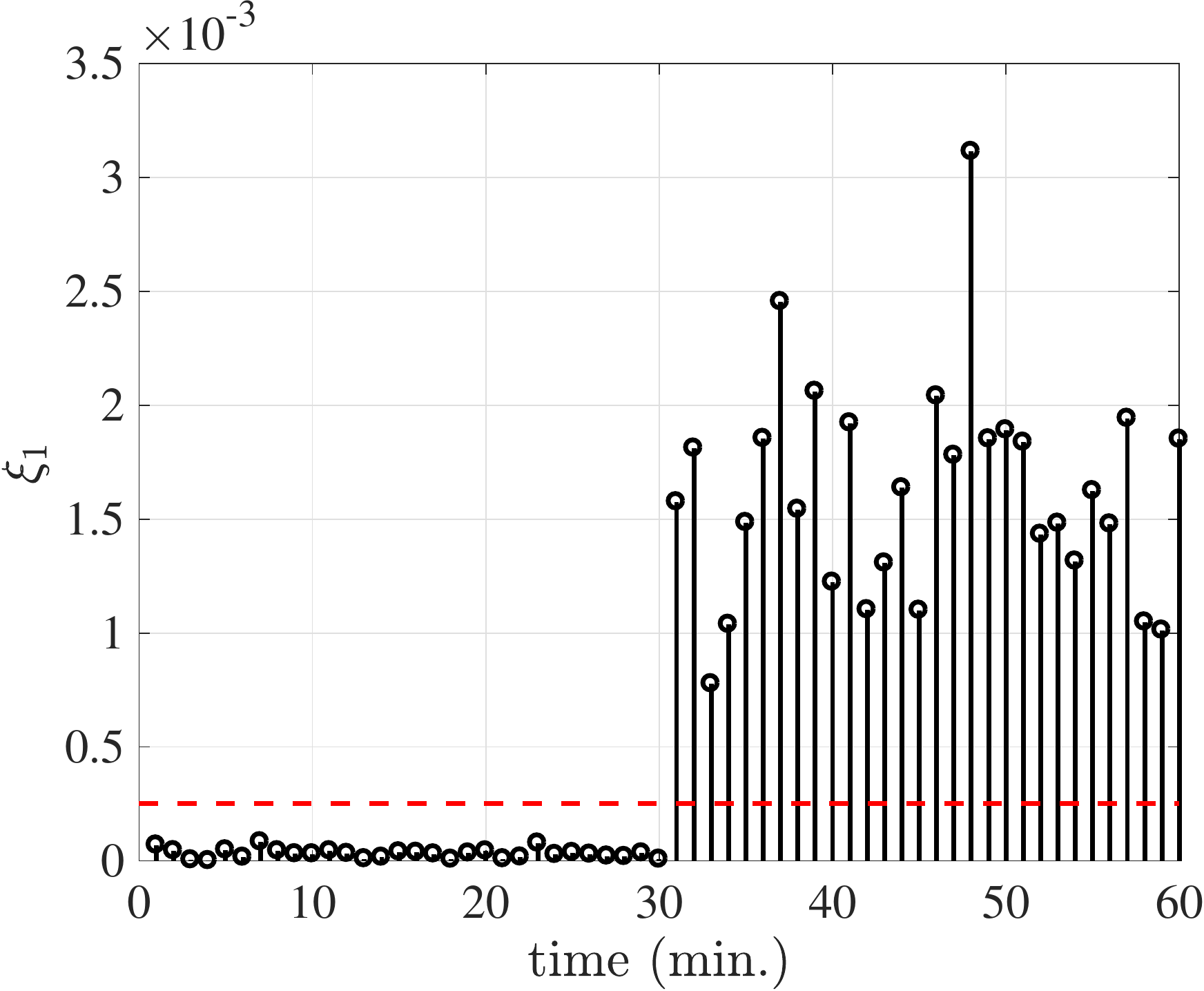}}
				\hfil
				\caption{The evolutions of indicator $\xi_1^{\text{j}}$ under the \textbf{noise-injection} attack on (a) the \textbf{frequency} measurement, and (b) the \textbf{tie flow} measurement, of \textbf{Area 1} starting at $30$ min.}
				\label{fig:noise_injection_detection_freq}
			\end{figure}

		In this section, we demonstrate the efficacy of the proposed approach for detection of noise-injection attacks. As mentioned in Sec. \ref{sub:attack_models}, additional noise is superimposed on the actual frequency measurement after the $30th$ minute, and it is chosen so that the frequency is still within the normal range. Fig. \ref{fig:noise_injection} shows the measurements of the frequency before and after the attack, and, again, we cannot notice any anomaly since the frequency is within the normal range all the time and no distinct feature ever appears after $30$ minutes.
		Using the proposed algorithm, the noise injection attack on the  frequency measurements (Fig. \ref{fig:noise_injection_detection_freq}) 
		is identified successfully between the 30th and 31st minutes. 

	\subsubsection{Detection of Destabilization Attack} 
	\label{subs:detection_of_destabilization_attack}
\begin{figure}[h!]
			\centering
			\subfloat[]{\includegraphics[width=1.7in]{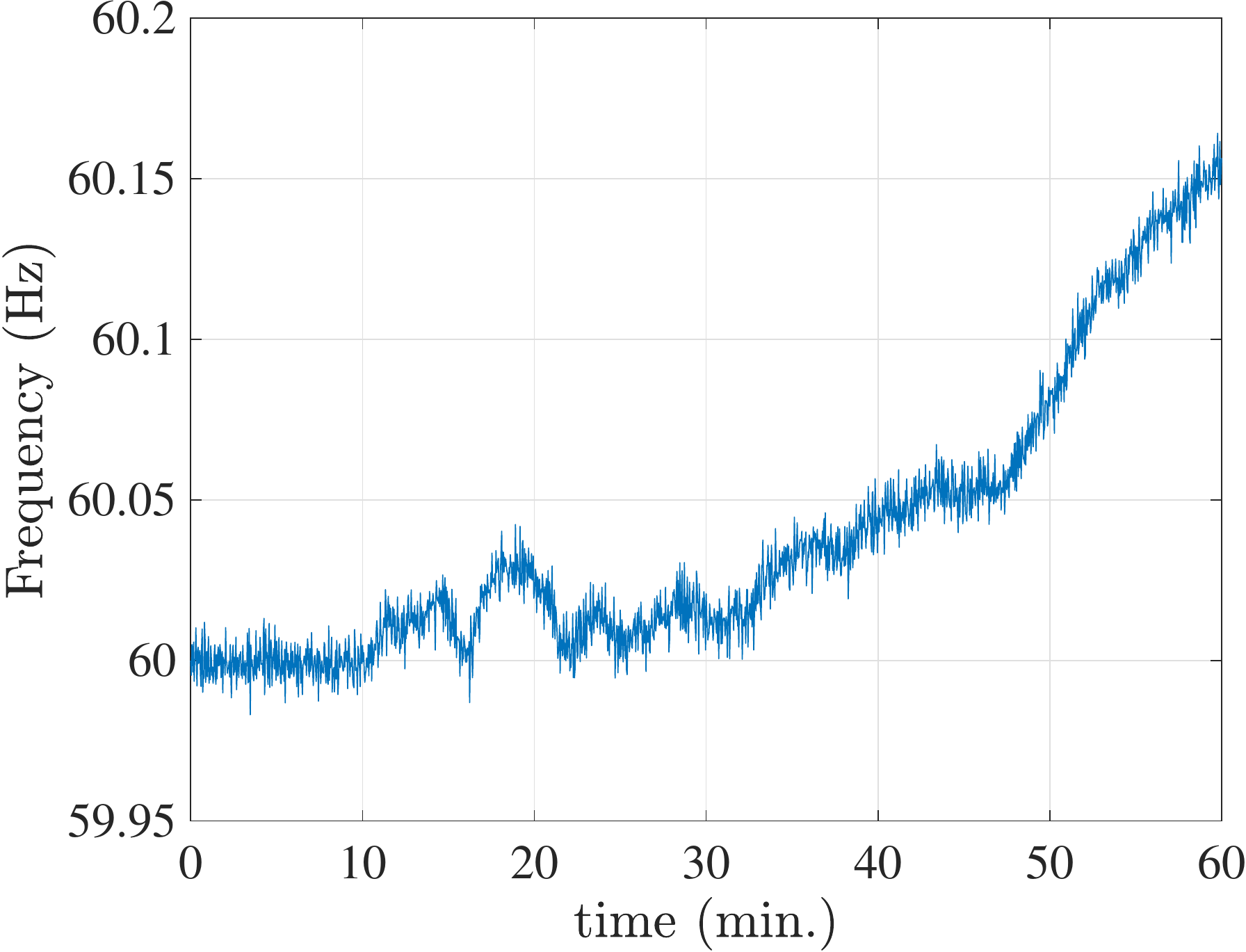}}
			\hfil
			\subfloat[]{\includegraphics[width=1.7in]{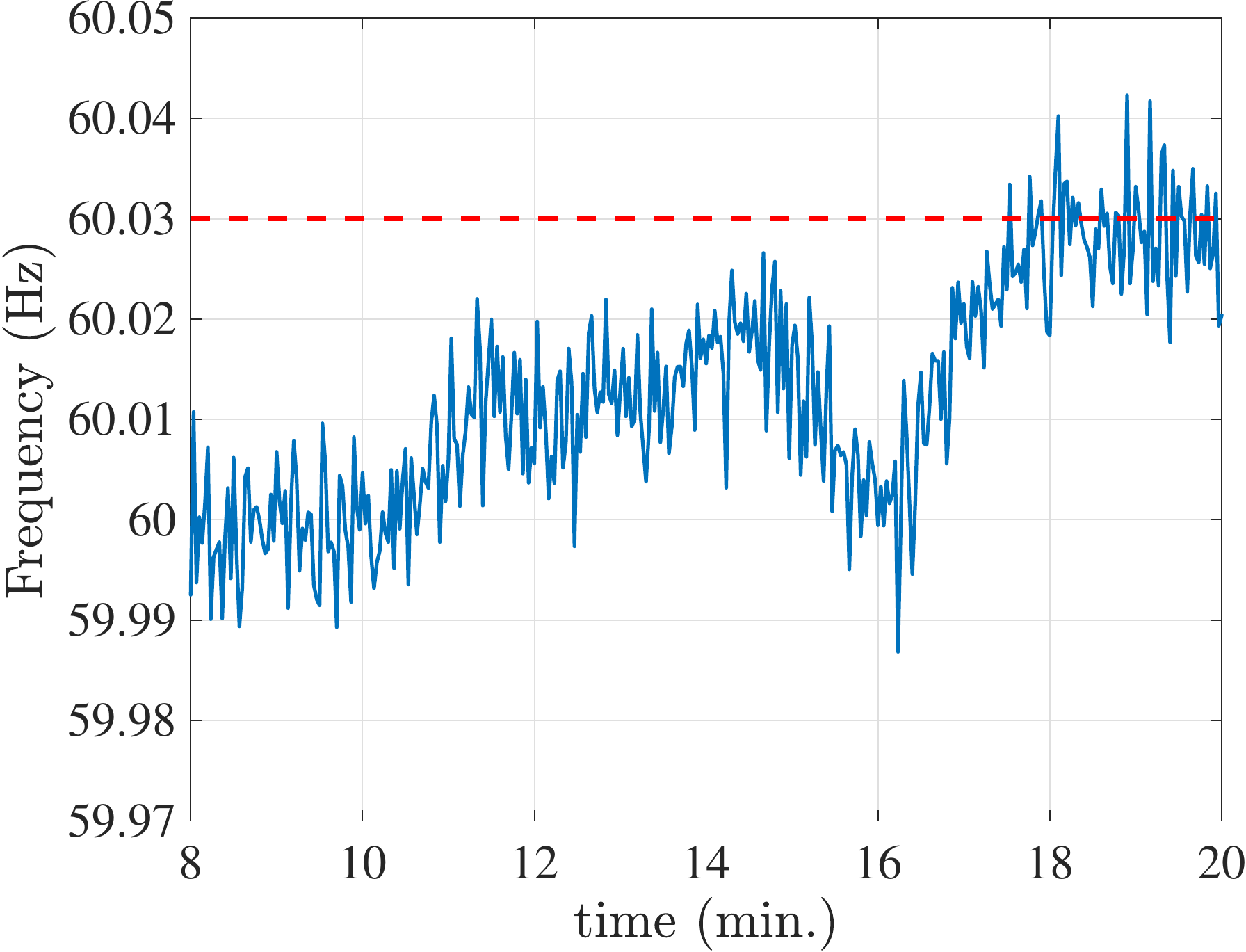}}
			\hfil
			\subfloat[]{\includegraphics[width=1.7in]{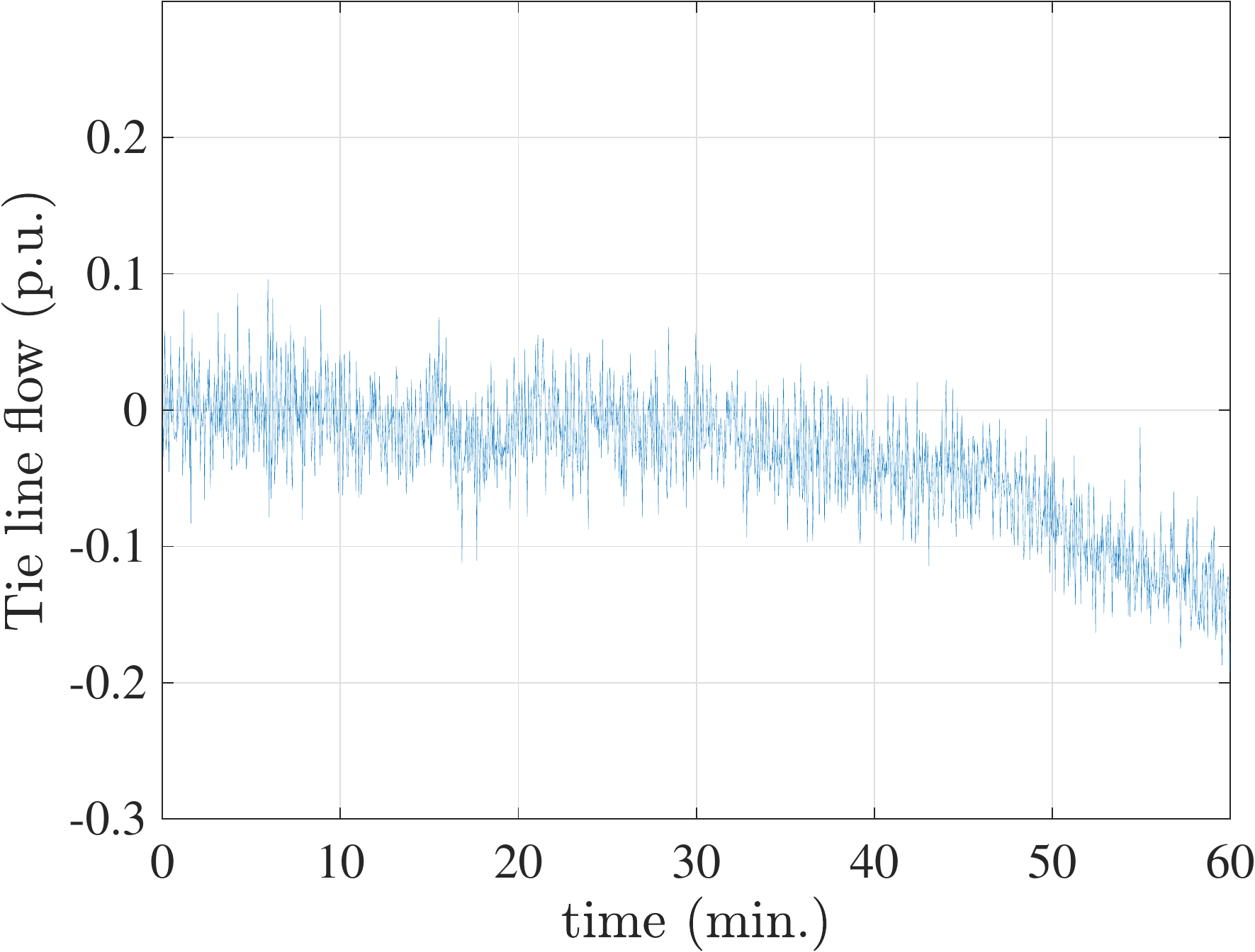}}
			\hfil
			\subfloat[]{\includegraphics[width=1.7in]{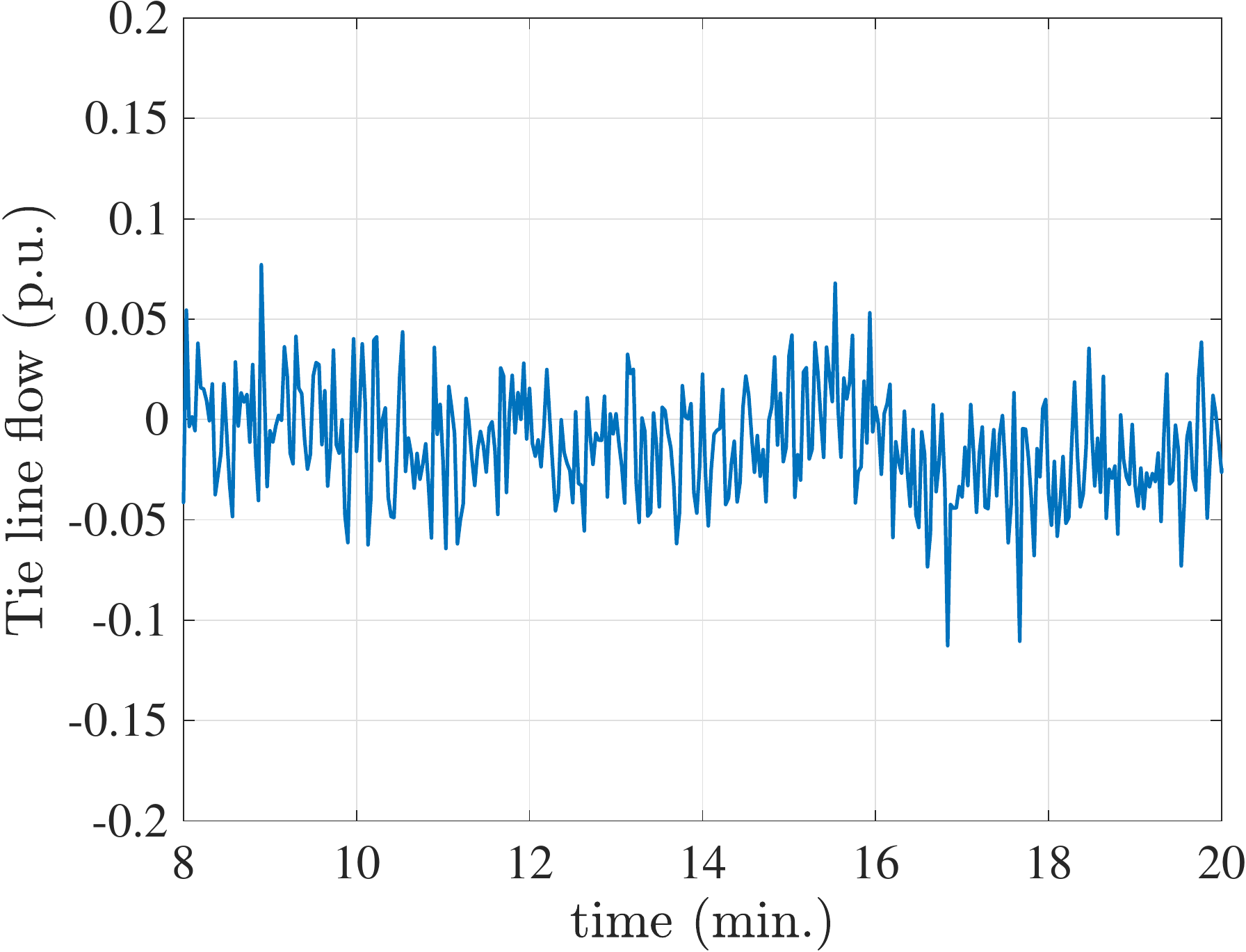}}
			\hfil
			\caption{Frequency measurement in Area $1$ from $0$ min to $60$ min (a) and its zoom-in frequency measurement (b), tie-line flow measurement in Area $1$ from $0$ min to $60$ min (c), and its zoom-in tie-line flow measurement (d) under the \textbf{destabilization} attack to the tie-line flow measurement of Area $1$ launched at $10$ min.}
			\label{fig: destable_measure}
		\end{figure}

	This section deals with securing the system from destabilization attacks. A destabilization attack is carried out on the tie-line flow measurements in Area $1$. As mentioned in Section \mbox{\ref{ssub:destabilization_attack}}, the output sequence of a malicious filter $M$ can be obtained via a simple tuning procedure, as follows. The adversaries may first force the sensor to report to the control center the scheduled tie-line flow plus a scaled version of actual flow deviation, i.e., \mbox{$p_{\text{sch}}+\lambda\Delta p$} with an arbitrary chosen \mbox{$\lambda$}, as opposed to the actual flow measurement \mbox{$p_{\text{sch}}+\Delta p$}. Then the attackers can gradually tune $\lambda$ such that the frequency exhibits unstable/oscillatory behavior. In the four-area system, the scalar \mbox{$\lambda$} is \mbox{$-0.89$}, and the attack starts at the \mbox{$10th$} minute. Based on the scaled flow measurement $\Delta p$, the control command is computed according to the AGC control law, and the load reference setpoint of Generators $1$, $5$ and $6$ are changed accordingly. As evident from Fig \ref{fig: destable_measure}(a), the closed-loop system is unstable and the frequency grows in an unbounded fashion. 

\begin{figure}[h!]
			\centering
			\includegraphics[width=3.5in]{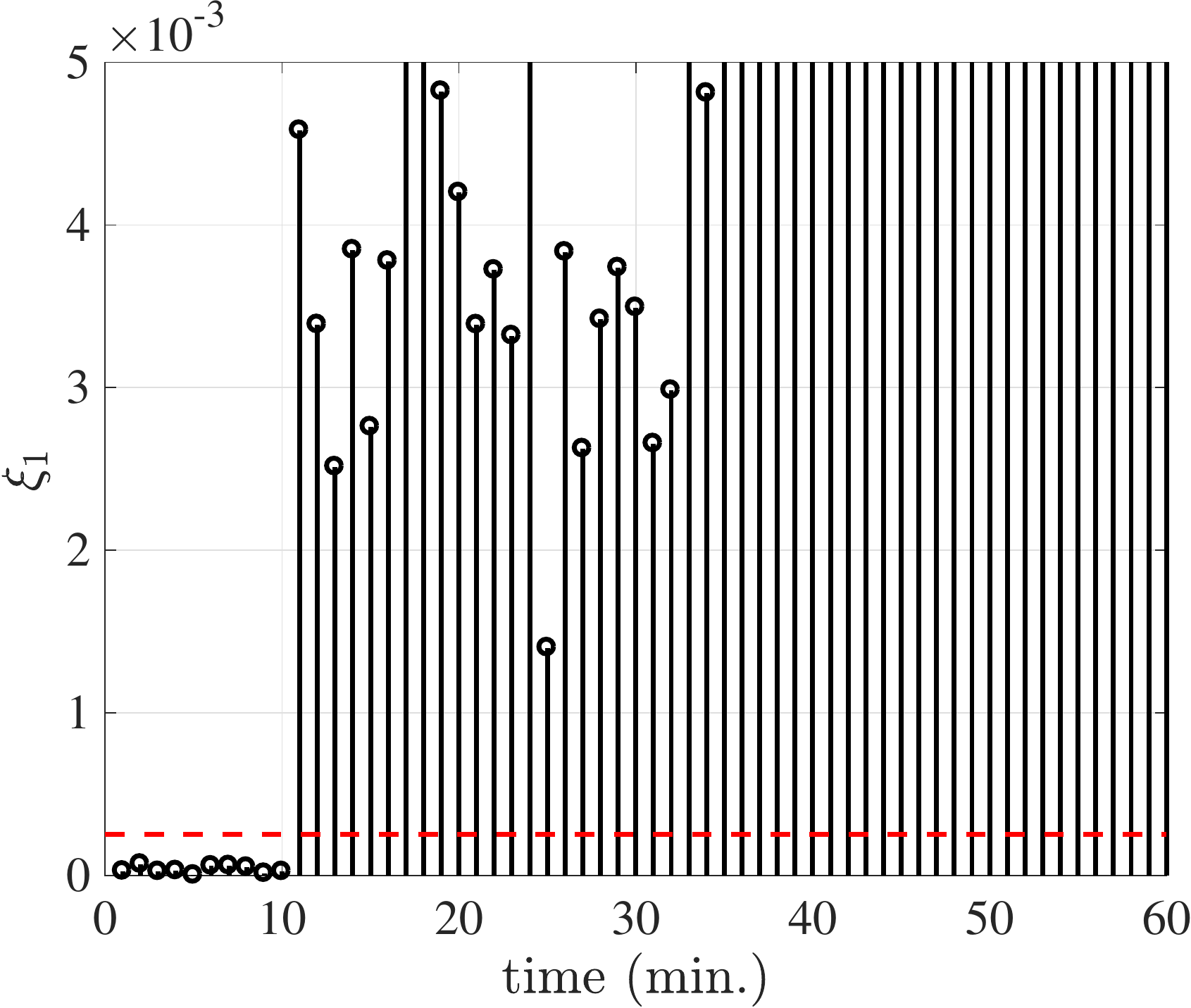}
			\caption{The evolution of indicator $\xi_1^{\text{j}}$ under the \textbf{destabilization} attack on the \textbf{tie flow} measurement of \textbf{Area $2$} starting at $10$ min.}
			\label{fig: distabilization_attack}
		\end{figure}

	Now we observe the process of destabilization attack from the perspective of the system operator. Suppose that the system operator keeps monitoring the reported frequency and tie-line flow measurements at the balancing authority of Area 1. Then, Fig. \ref{fig: destable_measure}(b) and Fig. \ref{fig: destable_measure}(d) are what the operator can observe from the $8th$ minute to the $20th$ minute. The operator might not realize the anomaly until around the $16th$ minute at which time several samples of frequency exceed the upper limit of the normal frequency range. However, the proposed approach can detect the destabilization attack between the $10th$ minute and $11th$ minutes, as we can see from Fig. \ref{fig: distabilization_attack}. 

\begin{figure}[h]
			\centering
			\subfloat[]{\includegraphics[width=1.7in]{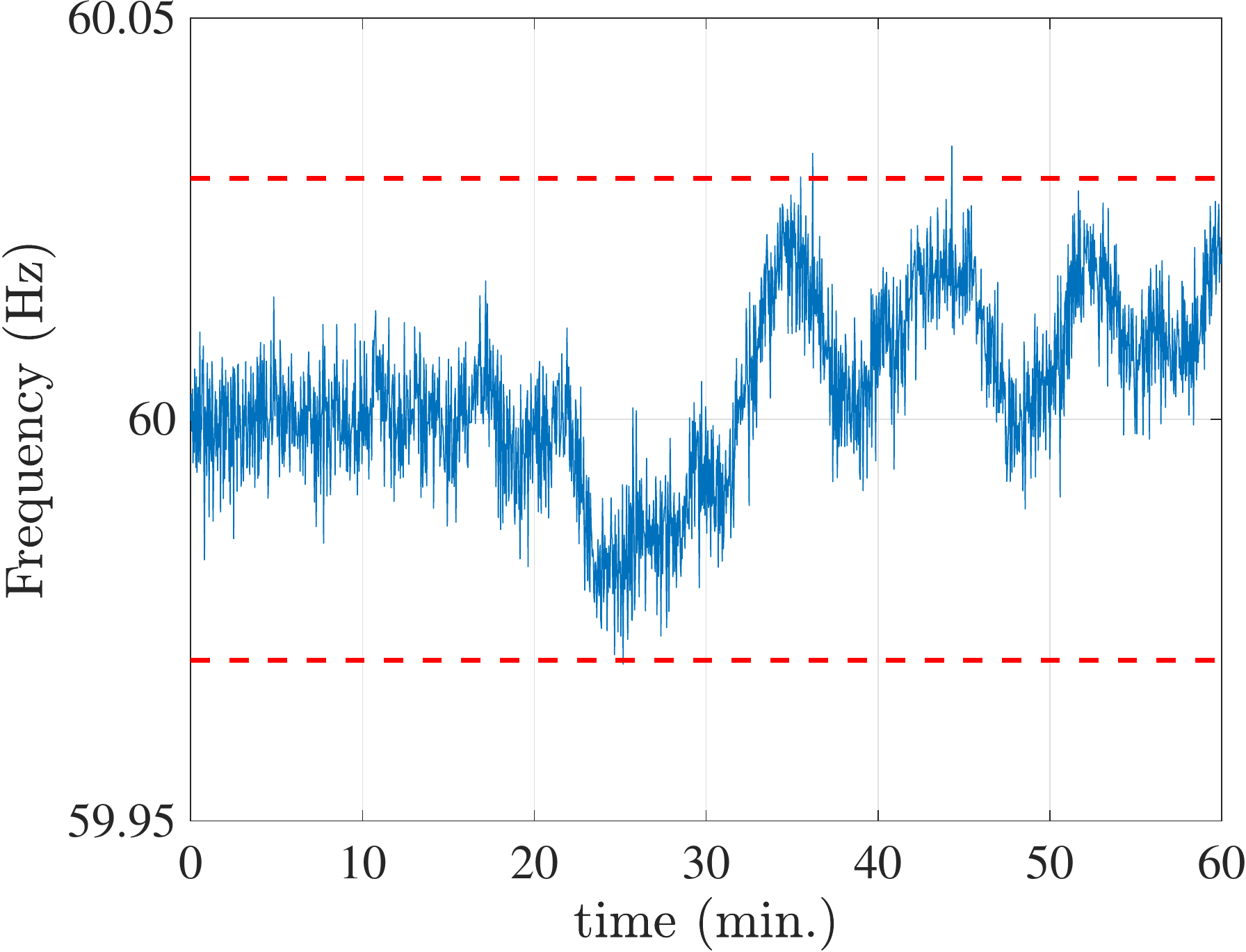}}
			\hfil
			\subfloat[]{\includegraphics[width=1.7in]{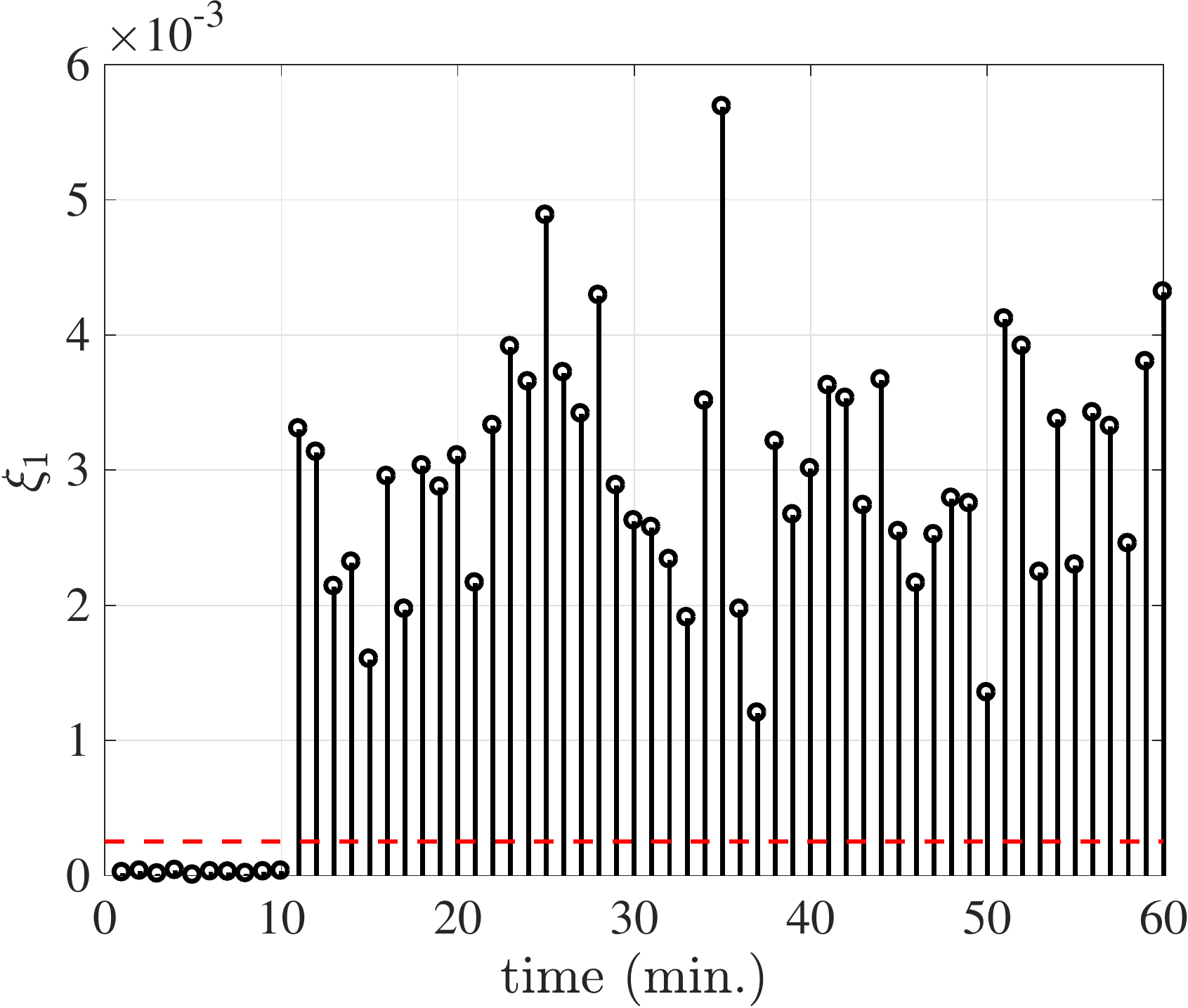}}
			\hfil
			\caption{Frequency measurement under \textbf{destabilization attack} to the \textbf{tie-line flow} measurement in Area 1 (a), and the evolution of corresponding $\xi_1^{\text{j}}$ (b).}
			\label{fig: distabilization_attack_oscillation}
		\end{figure}

	One might wonder if the ACE will always ultimately exceed its limits under a destabilization attack, in which
case the operator will notice it anyway, thereby rendering the proposed approach superfluous. The answer
is that there are sophisticated destabilization attacks where the ACE might not exhibit instability. Consider an attack template which is the same as earlier, except that $\lambda$ is set to $-0.84$. This results in the frequency measurement in Area $1$ shown in Fig. \ref{fig: distabilization_attack_oscillation}(a). It can be seen that though some frequency samples exceed the constraint occasionally, 
	these violations might be attributed to measurement error, and consequently be ignored by the operators since the frequency reverts to the normal range after several abnormal samples. In contrast the indicator signals under watermarking exhibit the consecutive spikes shown in Fig. \ref{fig: distabilization_attack_oscillation}(b) thereby detecting the attack on Area $1$. It can be seen that, in contrast to performing fine adjustments of the system frequency, the energy consumed by AGC drives the frequency to oscillate within a wider range compared to the frequency before the cyber attack.
\subsection{Performance Validation of the Proposed Algorithm on the NPCC 140-bus System} 
\label{sub:performance_validation_of_the_proposed_algorithm_on_the_NPCC_system}
	\subsubsection{NPCC 140-bus System Description and Parameter Setting of the Proposed Algorithm} 
	\label{ssub:npcc_140_bus_system_description_and_parameter_setting_of_the_proposed_algorithm}
	This benchmark system has 140 buses and 48 generators, and its raw parameters are available in the file named ``\mbox{\texttt{datanp48.m}}'' in PST \mbox{\cite{chow1992toolbox}}. In this paper, the NPCC 140-bus system is divided into two areas based on the geographical locations of buses \mbox{\cite{6913022}}, \mbox{\cite{8100989}}, and the bus indexes in each area are as reported in the Appendix. Accordingly, eight transmission lines are chosen as the tie lines connecting the two areas; they are Line 78-81 \footnote{Line 78-81 represents the transmission line from Bus 78 to 81.}, 76-77, 66-134, 67-138, 105-111, 105-106, 105-107, and 105-101. There are \mbox{$9$} generators in AGC loop, which are Generators 1, 2, 18, 19, 20, 21, 22, 23, and 24. The system matrices \mbox{$A$}, \mbox{$B$}, and \mbox{$C$} are extracted by PST. In Area 1, we add a discrete PI feedback loop, where both of the proportional gain constant and the integral gain are set to \mbox{$-0.0451$}. The variance parameter of the load deviations \mbox{$\sigma_{\text{L}}^2 = 0.001$} is chosen such that the frequency fluctuates within the normal range, i.e., \mbox{$60 \pm 0.03$} Hz \mbox{\cite{EPRI1234}} with high probability. The thresholds \mbox{$\eta_1=0.0045$} with \mbox{$\eta_1^{\infty}=6.3935\times 10^{-4}$} and \mbox{$\kappa'=7$}. The settings of \mbox{$\tau$}, \mbox{$\kappa_i$}, \mbox{$\sigma_f^2$}, \mbox{$Q'$}, \mbox{$T$}, \mbox{$\sigma_e$}, and SNR of deviation measurements of tie-line flow are the same as those in Section \mbox{\ref{sub:result_of_the_framework_for_the_four_area}}.

\begin{figure}[h!]
		\centering
		\includegraphics[width = 3.5in]{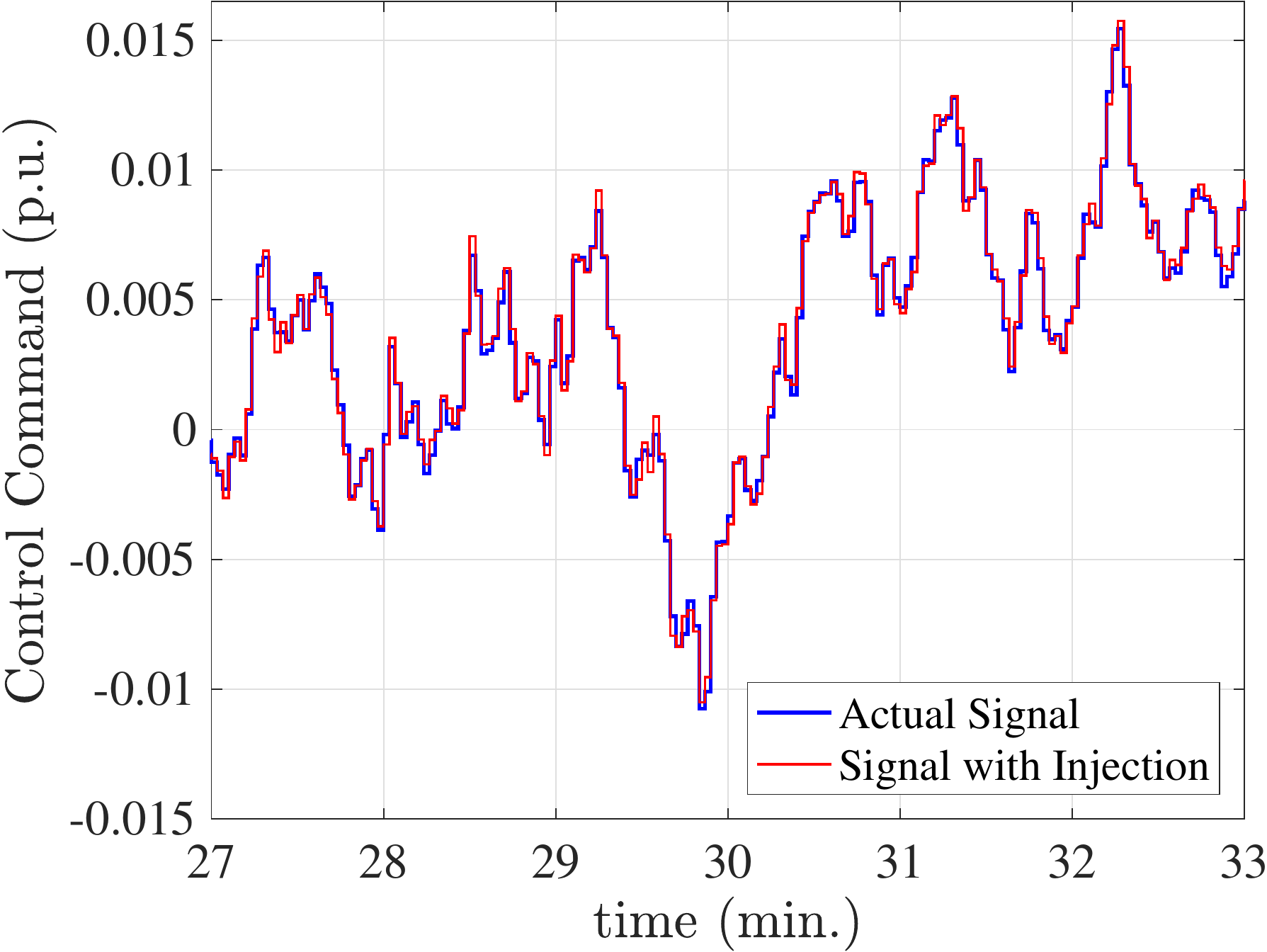}
		\caption{Control command comparison in the NPCC 140-bus power system.}
		\label{fig:Control_Command_NPCC}
	\end{figure}

	Again, we examine the impact of the private injection on the performance of the AGC in terms of frequency regulation. Figure \mbox{\ref{fig:Control_Command_NPCC}} records the control commands from AGC. It shows that the control command with the private injection does not deviate significantly from the control policy-specified input.
	\subsubsection{\mbox{Detection of Three Types of Cyber Attack}} 
	\label{ssub:detection_of_three_types_of_cyber_attack}
\begin{figure}[h!]
				\centering
				\subfloat[]{\includegraphics[width=1.7in]{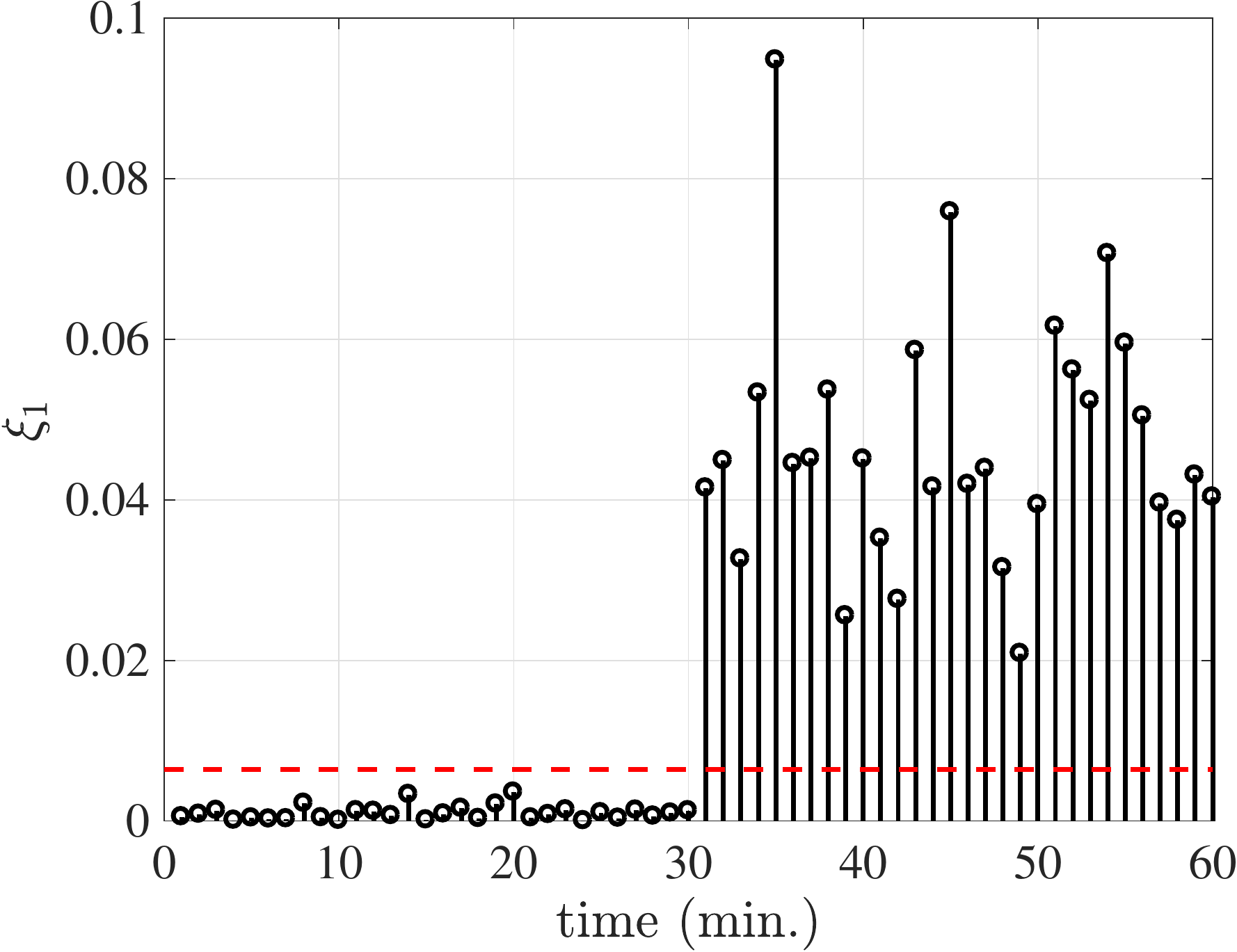}}
				\hfil
				\subfloat[]{\includegraphics[width=1.7in]{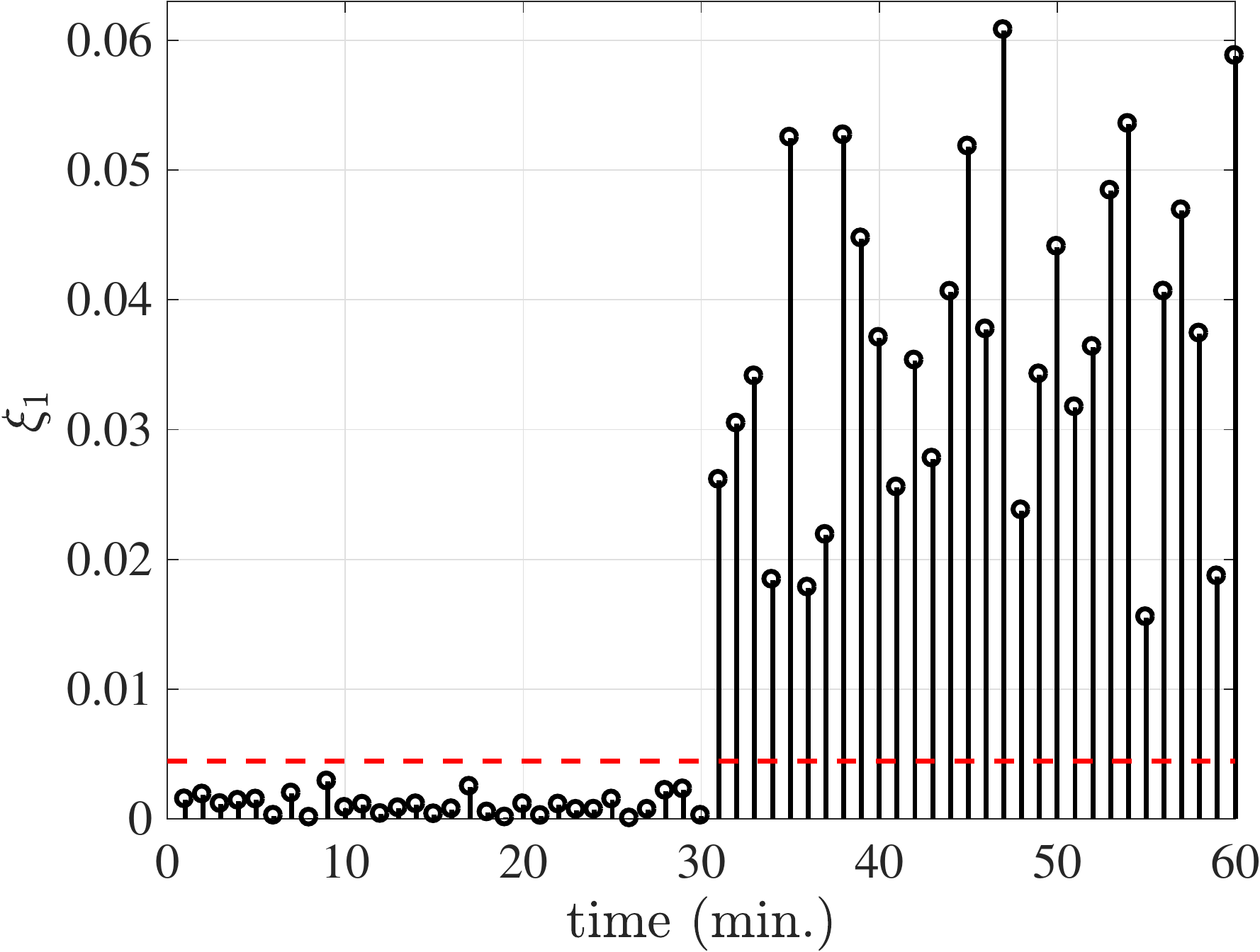}}
				\hfil
				\caption{The evolutions of indicator \mbox{$\xi_1^j$} under (a) the replay attack and (b) the injection attack on the NPCC 140-bus power system.}
				\label{fig:NPCC_Detection_first_two}
	\end{figure}

\begin{figure}[h!]
				\centering
				\subfloat[]{\includegraphics[width=1.7in]{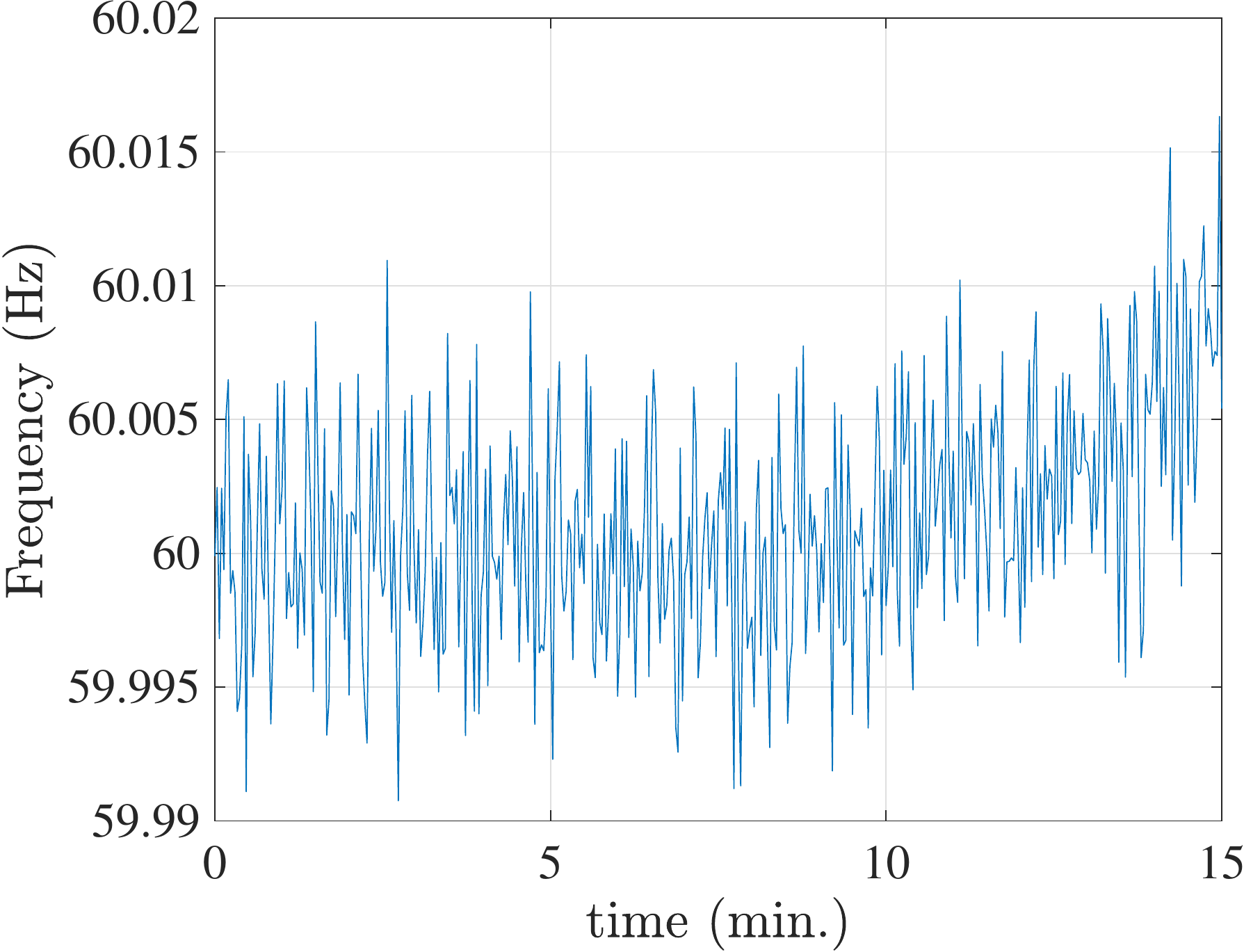}}
				\hfil
				\subfloat[]{\includegraphics[width=1.7in]{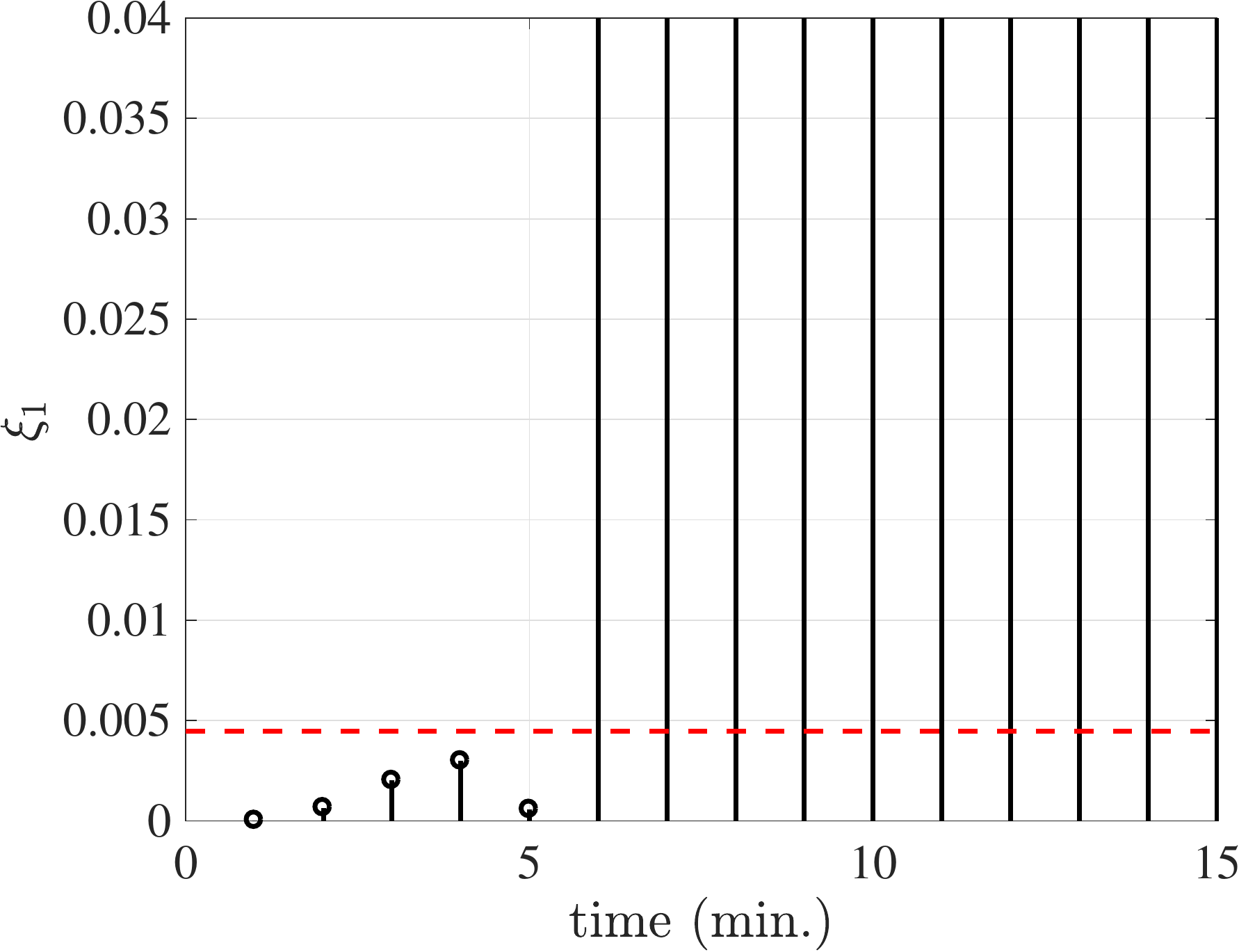}}
				\hfil
				\caption{(a) The time-domain frequency measurements under the destabilization attack; (b) the evolutions of indicator \mbox{$\xi_1^j$} under the destabilization attack.}
				\label{fig:NPCC_Detection_unstable}
	\end{figure}

	In this section, we demonstrate the efficacy of the proposed approach for detecting the three types of cyber attack defined in Sec. \mbox{\ref{sub:attack_models}}, through simulations on the NPCC 140-bus power system. We first validate the performance of the proposed algorithm in terms of detecting the replay attack and the noise-injection attack on the frequency measurement in the NPCC 140-bus system. Both types of cyber attack begin at \mbox{$30$} min. As shown in Fig. \mbox{\ref{fig:NPCC_Detection_first_two}}, both types of cyber attack are identified successfully between the \mbox{$30th$} and \mbox{$31st$} minutes. We next deal with securing the NPCC 140-bus system from the destabilization attacks. The destabilization attack on the flow measurement of Line 78-81 starts at the \mbox{$5th$} min, resulting in a growing trend of frequency deviation as shown in Fig. \mbox{\ref{fig:NPCC_Detection_unstable}}(a). The scalar \mbox{$\lambda$} defined in Section \mbox{\ref{subs:detection_of_destabilization_attack}} is \mbox{$-5$}. The evolution of \mbox{$\xi^j_1$} over time is presented in Fig. \mbox{\ref{fig:NPCC_Detection_unstable}}(b). It is observed that consecutive spikes exceed the threshold after the \mbox{$6th$} min, suggesting that the attack appears between the \mbox{$5th$} and \mbox{$6th$} minutes.

\subsection{Comparison with the Regression-based Approach} 
\label{sub:comparison_with_the_regression_based_approach}
In this section, we compare the dynamic-watermarking approach with the regression-based approach \mbox{\cite{7867825}} in the four-area system described in Section \mbox{\ref{sub:result_of_the_framework_for_the_four_area}}. In Reference \mbox{\cite{7867825}}, the cyber attacks on AGC are detected based on the following linear regression which characterizes the relationship between frequency (output) and load fluctuations (input), i.e.,
\begin{equation}
	\hat{\omega}(k) \approx \sum_{h=0}^{H-1}\boldsymbol{\alpha}_h\boldsymbol{u}_{\text{load}}(k-h).
	\label{eq:regression}
\end{equation}
Equation \mbox{\eqref{eq:regression}} assumes that the current frequency deviation \mbox{$\hat{\omega}(k)$} is a linear combination of the current load fluctuation vector \mbox{$\boldsymbol{u}_{\text{load}}(k)$} and the past load fluctuation vectors, i.e., \mbox{$\boldsymbol{u}_{\text{load}}(k-h)$} for \mbox{$h=1,2,\ldots,H-1$. $\boldsymbol{\alpha}_h$} is the combination coefficient vector, and the integer $H$ is the order of the linear regression, which is an adjustable factor. The state-space version of \mbox{\eqref{eq:regression}} can be identified by the MATLAB System Identification Toolbox \mbox{\cite{ljung1988system}}. The attack is detected by checking the discrepancies between the reported frequency measurement \mbox{$\omega(k)$} and its estimated value \mbox{$\hat{\omega}(k)$}. Hence, the indicator \mbox{$\gamma(k)$} in the regression-based framework is defined by \mbox{$\gamma(k):= \omega(k) - \hat{\omega}(k)$}. An alarm is triggered if
\begin{equation}
	|\gamma(k)|>\eta',
	\label{eq: regression_criteria}
\end{equation}
where \mbox{$\eta'$ is the maximal $|\gamma(k)|$} under the normal condition or during the training stage \mbox{\cite{7867825}}.

However, the regression-based approach may not detect the following cyber attacks on AGC. The linear regression \mbox{\eqref{eq:regression}} can be learned by a sophisticated adversary, based on the input-output measurements. Then, the threshold \mbox{$\eta'$} can be approximately estimated. Finally, the actual measurement can be replaced by the following malicious measurement sequence \mbox{$\omega_a$} without being detected by the criteria \mbox{\eqref{eq: regression_criteria}}:
\begin{equation}
	\omega_{\text{a}} = \hat{\omega} -\eta'.
	\label{eq: smart_attack}
\end{equation}

\begin{figure}[h!]
				\centering
				\subfloat[]{\includegraphics[width=1.7in]{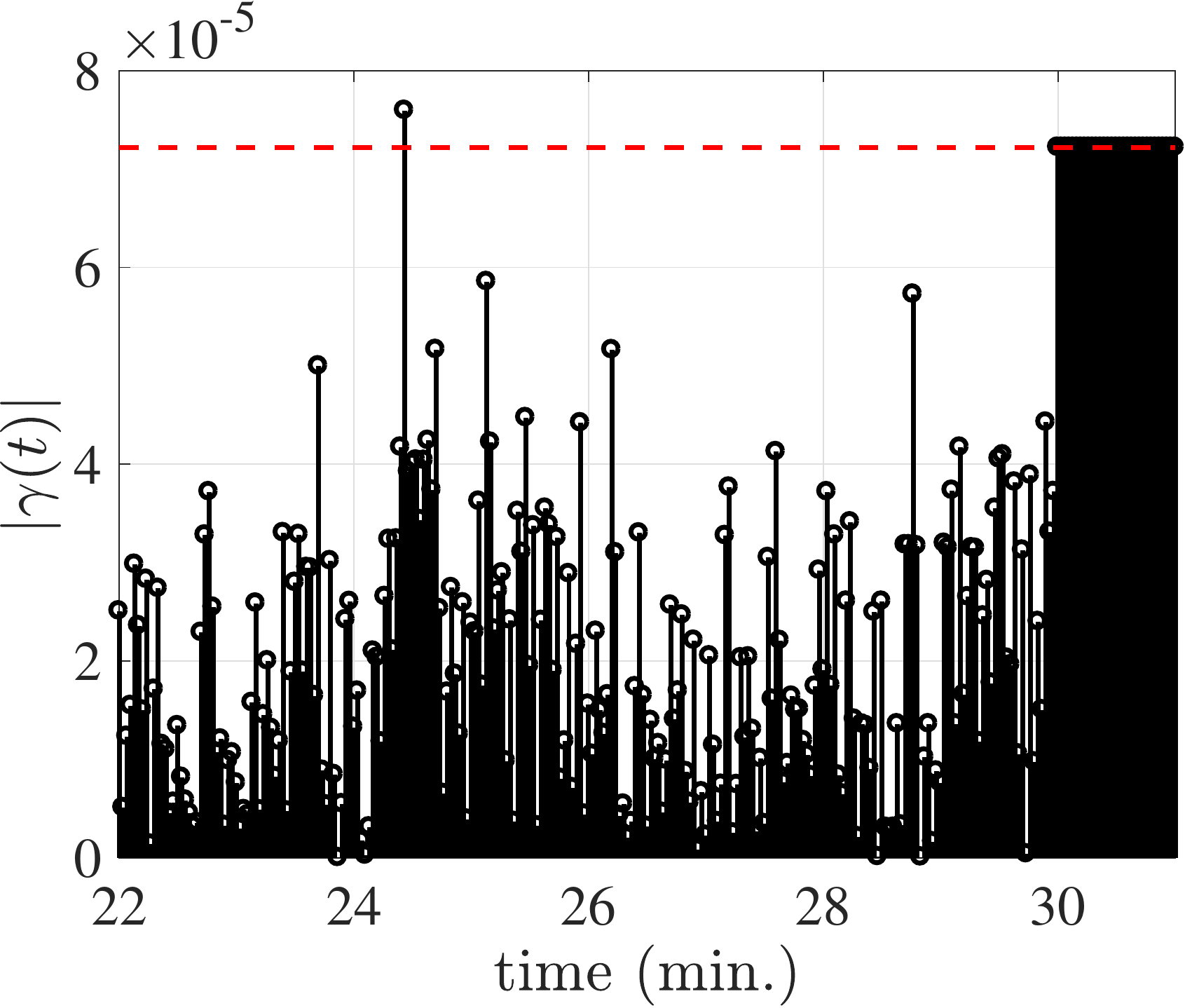}}
				\hfil
				\subfloat[]{\includegraphics[width=1.7in]{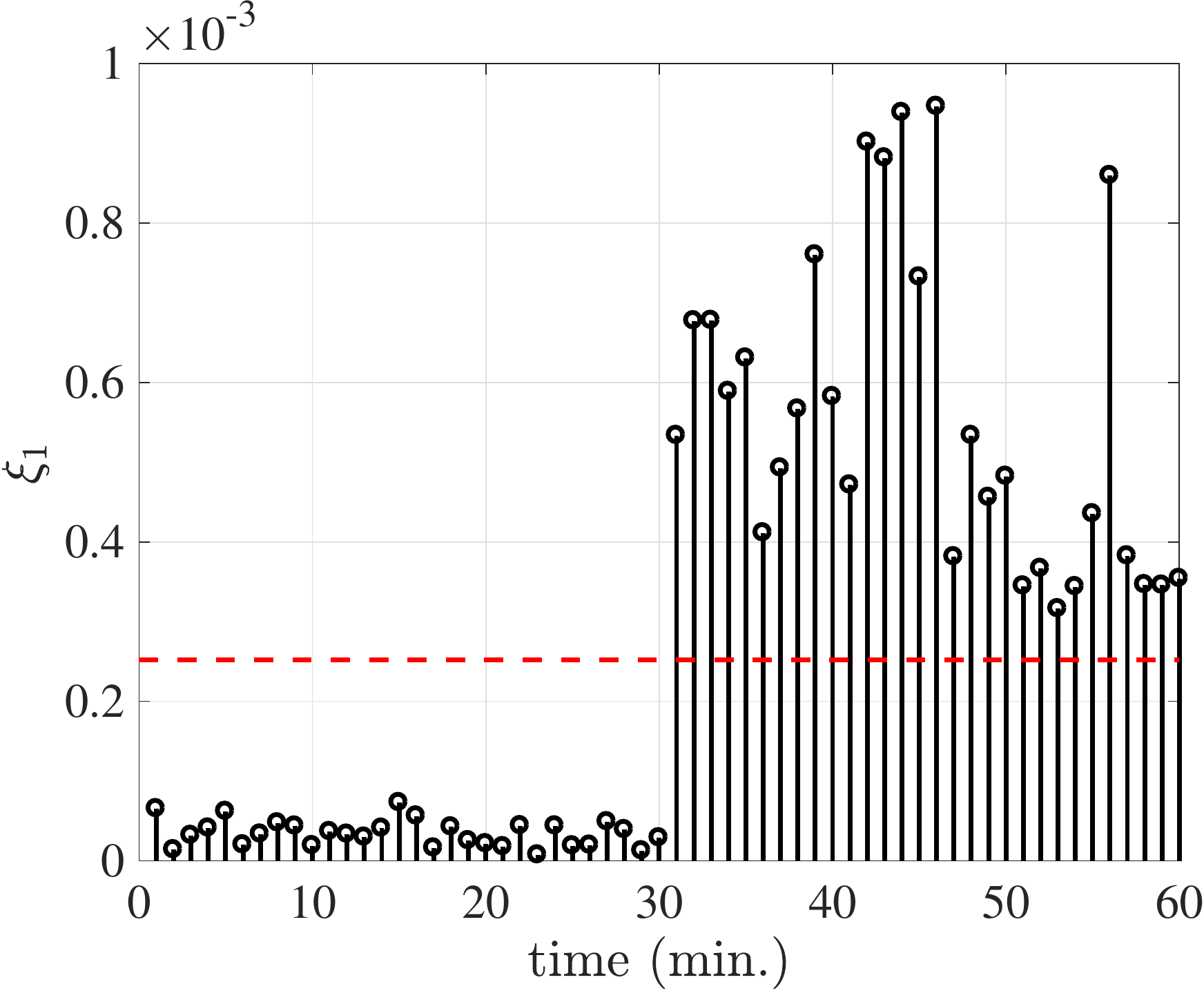}}
				\hfil
				\caption{The evolutions of (a) \mbox{$|\gamma_(t)|$} and (b) \mbox{$\xi_i^j$} over time under the attack defined in \mbox{\eqref{eq: smart_attack}}.}
				\label{fig:Smart_Attack}
			\end{figure}

Next, we test the performance of the proposed algorithm in terms of detecting the attack defined in \mbox{\eqref{eq: smart_attack}}. The attack with \mbox{$\eta'=9.024\cross10^{-5}$} starts at \mbox{$30$} min in the four-area system. Fig. \mbox{\ref{fig:Smart_Attack}}(a) presents the evolution of the \mbox{$|\gamma(t)|$} defined in the regression-based approach. It can be seen that \mbox{$|\gamma(t)|$} does not exceed the threshold \mbox{$\eta'$} after the \mbox{$30th$} minute, although it keeps being close to \mbox{$\eta'$}. In contrast, the indicator under the proposed method exceeds a predefined threshold consecutively after the \mbox{$30th$} minute, suggesting that the attack defined in \mbox{\eqref{eq: smart_attack}} can still be identified successfully, as shown in Fig. \mbox{\ref{fig:Smart_Attack}}(b). Note that, although the regression-based approaches are not guaranteed to detect any cyber attacks, it can serve as a screening tool for the proposed framework.

\subsection{Robustness Test} 
\label{sub:robustness_test}
Due to the effect of deadband in generation units, some generators might not be responsive to small change in setpoints, and these generators are termed as non-responsive generators (NRGs). The number of non-responsive generators in AGC may impact the performance of the proposed framework. In order to investigate such an impact, we first define a performance indicator $\theta$. In the context of Section \mbox{\ref{sub:performance_validation_of_the_proposed_algorithm_on_the_NPCC_system}}, where the cyber attack (replay/injection) starts from the \mbox{$30th$} min, the performance indicator \mbox{$\theta$} can be defined as follows:
\begin{equation}
	\theta = \frac{\min_p {\xi_1^p}}{\max_q {\xi_1^q}} \quad \forall p\in\{31, 32,\ldots,60\} \land q\in\{1,2,\ldots,30\},
\end{equation}
where the numerator suggests the minimal value of \mbox{$\xi_1^j$} under the attack, while the denominator is the maximal value of \mbox{$\xi_1^j$} under the normal condition. If the ratio \mbox{$\theta> 1$, $\xi_1^j$} under the attack can be linearly separated from that under the normal condition by setting a threshold, i.e, the attack can be detected by the proposed method.

In Section \mbox{\ref{sub:performance_validation_of_the_proposed_algorithm_on_the_NPCC_system}}, we assume that all \mbox{$9$} generators in the AGC loop are responsive to small changes in their setpoints. Here, we increase the number of the NRGs from \mbox{$0$} to \mbox{$5$}, and compute the corresponding performance indicators \mbox{$\theta$} under the replay attack and the noise-injection attack. The results are presented in Table \mbox{\ref{tab:responsive_test}}. It is seen that both \mbox{$\theta_R$} and \mbox{$\theta_I$ a}re greater than \mbox{$1$} under all scenarios, suggesting that the replay attack and the noise injection attack can still be detected, even though some non-responsive generators exist.

\begin{table}[h]
	\caption{The Impact of Number of Responsive Generators (\mbox{$\theta_R$}: \mbox{$\theta$} under the \mbox{\underline{R}}eplay Attack; \mbox{$\theta_I$}: \mbox{$\theta$} under the \mbox{\underline{I}}njection Attack)}
	\label{tab:responsive_test}
	\centering
\begin{tabular}{c|l|c|c}
	\hline

	\hline
	\% of NRGs & NRG Index & $\theta_{\text{R}}$ & $\theta_{\text{I}}$ \\
	
	\hline
		$0/9$ & N/A & 6.5039 &7.1692\\
	\hline
		$1/9$ & $24$ & 6.4115 &7.0572\\
	\hline
		$2/9$ & $23, 24$ & 6.3088 &6.9370\\
	\hline
		$3/9$ & $22, 23, 24$ & 6.1079 &6.6742\\
	\hline
		$4/9$ & $21, 22, 23, 24$ & 6.1480 &6.7020\\
	\hline
		$5/9$ & $20, 21, 22, 23, 24$ & 6.0231 &6.5594\\
	\hline

	\hline
	\end{tabular}
\end{table}

\section{Conclusion} 
\label{sec:conclusion}
	In this paper, an online framework to detect cyber attacks on AGC is proposed. In the proposed defense framework, a theoretically rigorous attack detection algorithm based on dynamic watermarking is developed to detect sophisticated adversaries equipped with extensive and even complete knowledge of the physical and statistical models of the power system. The proposed framework needs no hardware update of the generation units.
	The efficacy of the proposed framework is demonstrated in a four-area synthetic power system and a 140-bus power system. 
	Future work will investigate the scaling up of the proposed method to larger-scale power systems.

\appendices
\section{Determining the Threshold Using the Neyman-Pearson criterion} 
\label{sec:determining_the_threshold_using_the_neyman_pearson_criterion}
The detection thresholds can also be determined using the Neyman-Pearson criterion based on the maximum tolerable false alarm rate $\theta_0$. Such a test is developed below.

We first note that the innovations process \mbox{$\{\zeta\}$} in the paper is distributed according to \mbox{$\zeta_k\sim\mathcal{N}(0,C_{di}PC_{di}^T+R)$} and i.i.d. across time (notation used is same as that in the paper).  Therefore, it can be shown that \mbox{$\zeta^T(k)\zeta(k)=\zeta_1^2(k)+\hdots+\zeta_n^2(k)$} is distributed according to the distribution of \mbox{$\sum_{i=1}^n\lambda_iZ_i^2,$} where \mbox{$\lambda_1,\hdots,\lambda_n$} are the eigenvalues of the covariance matrix \mbox{$C_{di}PC_{di}^T+R,$} and \mbox{$Z_1,\hdots,Z_n$} are i.i.d. Gaussian random variables with zero mean and unit variance. The distribution of \mbox{$\zeta^T(k)\zeta(k)$}, therefore, is that of a weighted linear combination of \mbox{$n$} independent chi-squared random variables with one degree of freedom each, and the distribution of the test statistic \mbox{$\xi_1^j$} in the paper is obtained by a \mbox{$T-$}fold convolution of the above distribution (since it is the sum of \mbox{$T$} i.i.d. random variables) followed by a translation to make the resulting mean zero. Denote this distribution by \mbox{$f_0,$} and let \mbox{$H_0$}, the null hypothesis, be the case when the system is not under attack. Therefore,
\begin{subequations}
\begin{align*}
H_0 : \xi_1^j \sim f_0,\\
H_1 : \xi_1^j \sim f_1,
\end{align*}
\end{subequations}
where \mbox{$f_1\neq f_0$} is any arbitrary distribution.

Consider the threshold test
 	\[
 		f_0(\xi_1^j)\underset{H_0}{\overset{H_1}{\gtrless}} \gamma_0
 	\]
 where \mbox{$\gamma_0$} is the detection threshold, a value that is chosen in the Neyman-Pearson test depending on \mbox{$\theta_0$}, the maximum tolerable false alarm rate. The false alarm rate for the test is given by $$\mathbb{P}(H_1|H_0)=\int_{x:f_0(x)<\gamma} f_0(x)dx,$$ which when viewed as a function of \mbox{$\gamma_0,$} is monotonically increasing. The optimal threshold \mbox{$\gamma_0^*$} is determined as
\begin{align*}
arg\max\int_{x:f_0(x)<\gamma_0} f_0(x)dx\\
\textit{s.t.} \;\;\;\;\int_{x:f_0(x)<\gamma_0} f_0(x)dx < \theta_0. 
\end{align*}

Since the function \mbox{$f_0$} is known, the threshold \mbox{$\eta_1$} in the paper can be determined from \mbox{$\gamma_0^*$}. A similar procedure is followed to determine the value of the threshold \mbox{$\eta_2.$}

There are two notions of detection performance in this context: (i) detection delay, and (ii) false alarm rate. These are conflicting objectives, and the problem of designing ``optimal" detection schemes in our context is a special instance of this more general ``quickest change detection" problem, and in what follows, we outline how exactly our problem maps to the latter. Further analysis of detection performance, such as optimal trade-offs between the detection delay and the false alarm rate, or optimal detection mechanisms for various objective functions, are well-studied problems in the quickest change detection literature (see \mbox{\cite{tartakovsky2005general}}, \mbox{\cite{lorden1971}}, \mbox{\cite{pollak1985}}, \mbox{\cite{doi:10.1137/1108002}}, \mbox{\cite{VEERAVALLI2014209}} and references therein). 

	The quickest change detection problem has three ingredients \mbox{\cite{VEERAVALLI2014209}}: (i) a stochastic process that is under observation, (ii) a random time \mbox{$\tau$} at which the statistics of the process changes from one distribution to another, and (iii) a detection algorithm that observes the stochastic process and declares at each time \mbox{$t$} whether \mbox{$t\geq\tau$} (change has occurred) or \mbox{$t<\tau$} (change has not yet occurred). This problem has a long history with applications in various fields such as manufacturing systems, quality control, network security, econometrics, etc. 

	In the context of our paper, the innovations process corresponding to the state estimate process \mbox{$\{\widehat{x}_{di}(k|k)\}$} is the stochastic process that is under observation or test, and the time at which the statistics of the process changes is the time at which the attack is initiated. As prior works show \mbox{\cite{7738534}}, \mbox{\cite{7945354}}, in the presence of dynamic watermarking, the adversary cannot introduce any significant distortion without causing a change in the statistics of the stochastic process under test, no matter what attack strategy it chooses to employ. Thus, formulating our problem as one of quickest change detection provides us with a mature framework for developing different statistical tests that are optimal for various objective functions.

\section{Parameters of the Four-area Power System and Bus Information of Each Area in the NPCC 140-bus System}
The four-area system used in Section \mbox{\ref{sub:result_of_the_framework_for_the_four_area}} is modified from the benchmark two-area system \mbox{\cite{kundur1994power}}, \mbox{\cite{chow1992toolbox}}. The information on the parameters of buses and transmission lines of the four-area system is reported in \cite{bus_branch_info}. In the four-area system, all generators associated with their exciters, governors and power system stabilizers (PSS) are the same, and their parameters can be found in ``\mbox{\texttt{d2asbegp.m}}'' in PST \mbox{\cite{chow1992toolbox}}.

The NPCC 140 bus system is divided into two areas in this paper. The buses in the Area 1 are 1, 2, 3, 4, 5, 6, 7, 8, 9, 10, 11, 12, 13, 14, 15, 16, 17, 18, 19, 20, 21, 22, 23, 24, 25, 26, 27, 28, 29, 30, 31, 32, 33, 34, 35, 36, 37, 38, 39, 40, 41, 42, 43, 44, 45, 46, 47, 48, 48, 49, 50, 51, 52, 53, 54, 56, 57, 58, 59, 60, 61, 62, 63, 64, 64, 65, 66, 67, 68, 70, 71, 72, 73, 74, 77, 78, 79, 80, 91, 92, 93, 94, 95, 96, 97, 99, 100, 101, 102, 103, 104, 106, 107, 108, 109, 110, 111, and 140. The rest of buses are in the Area 2.


\bibliographystyle{IEEEtran}
\bibliography{IEEEabrv,J2ref}

		




\ifCLASSOPTIONcaptionsoff
  \newpage
\fi

\begin{IEEEbiography}[{\includegraphics[width=1in,height=1.25in,clip,keepaspectratio]{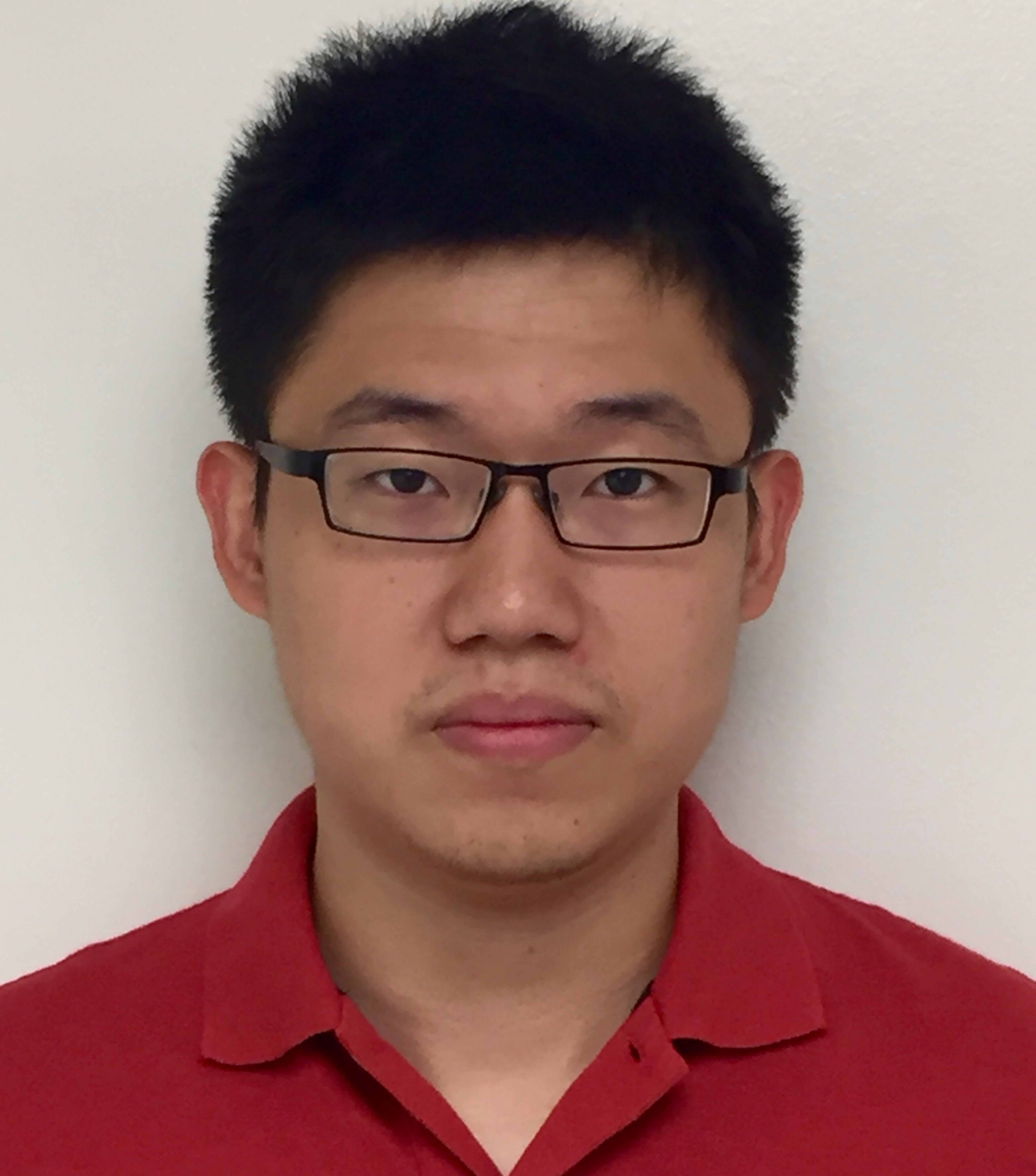}}]{Tong Huang}
(S'16) received the B.E. degree in Electric Power Engineering and its Automation from
North China Electric Power University, Baoding, China, in 2013 and the M.S. degree in Electrical
Engineering from Texas A\&M University, College Station, TX, USA, in 2017, where he is currently
working toward the Ph.D. degree. His research interests include power system stability, security, and wide-area monitoring and control systems.
\end{IEEEbiography}

\begin{IEEEbiography}[{\includegraphics[width=1in,height=1.25in,clip,keepaspectratio]{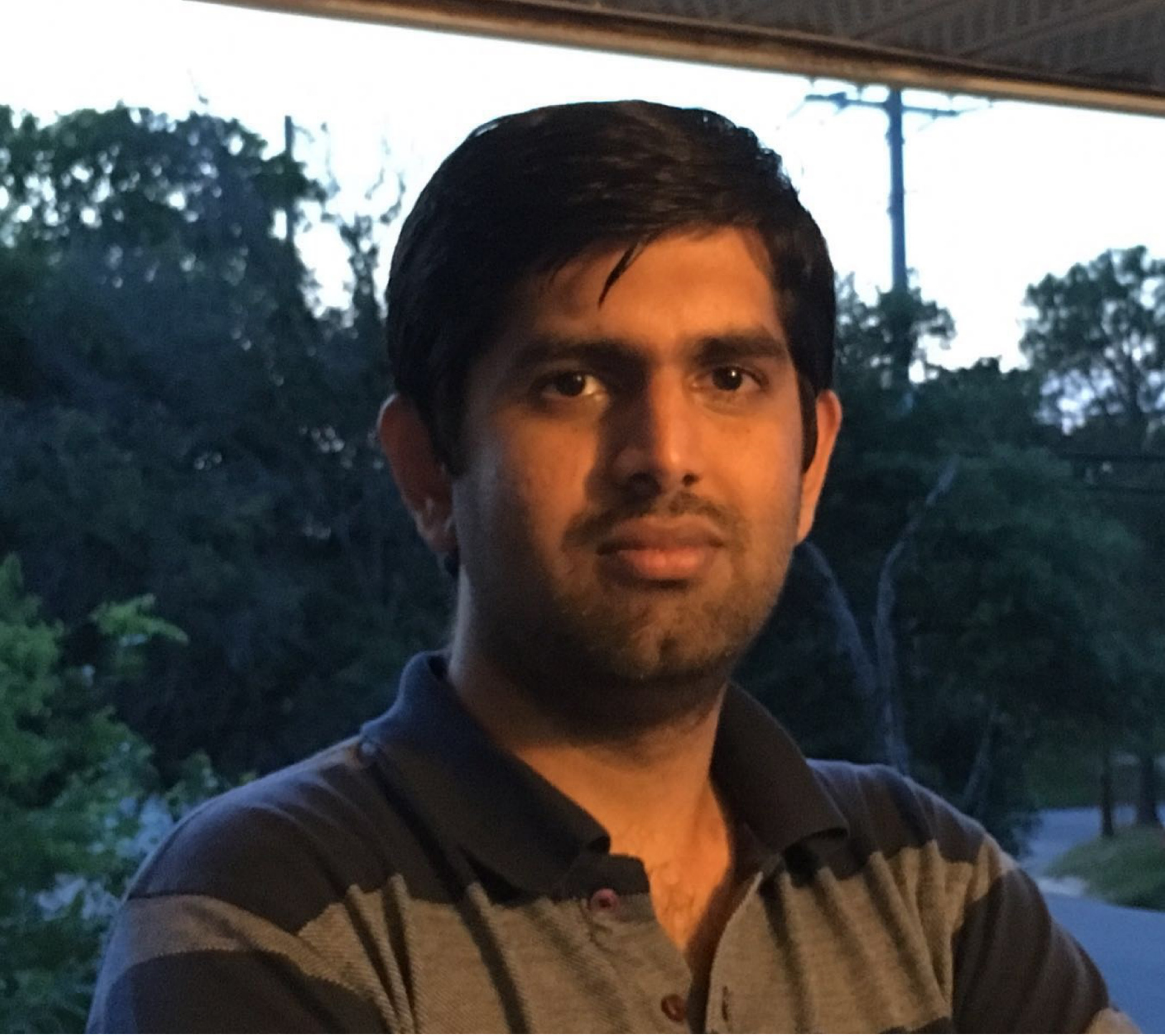}}]{Bharadwaj Satchidanandan} is a doctoral student in the Electrical and Computer Engineering department at Texas A\&M University, College Station, TX. Prior to this, he completed his Master’s from Indian Institute of Technology Madras, where he worked on Wireless Communications. Between May ’15 and August ’15, he interned at Intel Labs, Santa Clara, CA, where he worked on interference cancellation algorithms for next-generation wireless networks. His research interests include cyberphysical systems, power systems, security, database privacy, communications, control, and signal processing.
\end{IEEEbiography}

\begin{IEEEbiography}[{\includegraphics[width=1in,height=1.25in,clip,keepaspectratio]{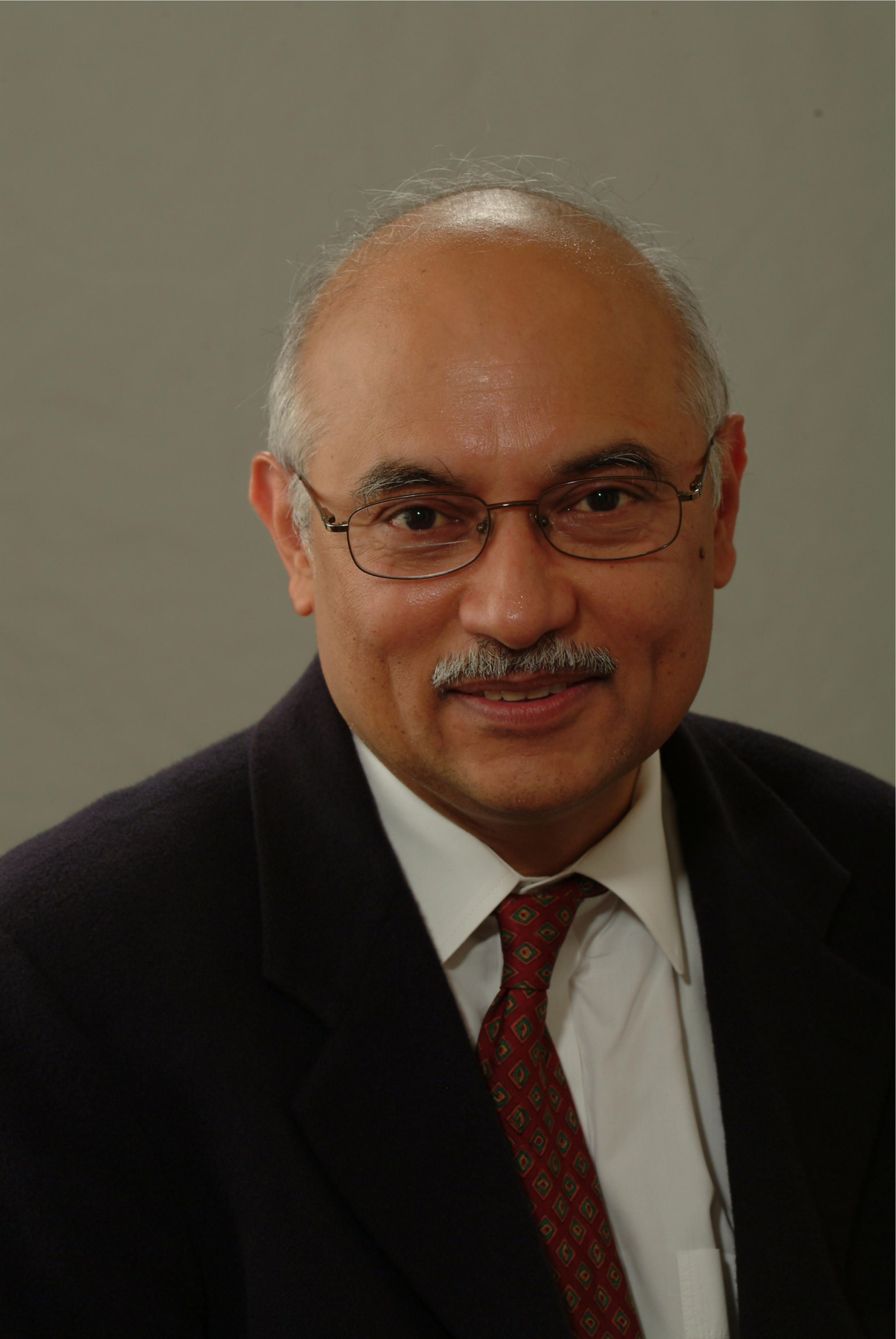}}]{P. R. Kumar} B. Tech. (IIT Madras, 73), D.Sc.(Washington University, St. Louis, 77), was a faculty member at UMBC (1977-84) and Univ. of Illinois, Urbana-Champaign (1985-2011). He is currently at Texas A\&M University. His current research is focused on stochastic systems, energy systems, wireless networks, security, automated transportation, and cyberphysical systems. He is a member of the US National Academy of Engineering and The World Academy of Sciences. He was awarded a Doctor Honoris Causa by ETH, Zurich. He has Award for Control Systems, the Donald P. Eckman received the IEEE Field
Award of the AACC, Fred W. Ellersick Prize of the IEEE Communications Society, the Outstanding Contribution Award of ACM SIGMOBILE, the INFOCOM Achievement Award, and the SIGMOBILE Test-of-Time Paper Award. He is a Fellow of IEEE and ACM Fellow. He was Leader of the Guest Chair Professor Group on Wireless Communication and Networking at Tsinghua University, is a D. J. Gandhi Distinguished Visiting Professor at IIT Bombay, and an Honorary Professor at IIT Hyderabad. He was awarded the Distinguished Alumnus Award from IIT Madras, the Alumni Achievement Award from Washington Univ., and the Daniel Drucker Eminent Faculty Award from the College of Engineering at the Univ. of Illinois.
\end{IEEEbiography}

\begin{IEEEbiography}[{\includegraphics[width=1in,height=1.25in,clip,keepaspectratio]{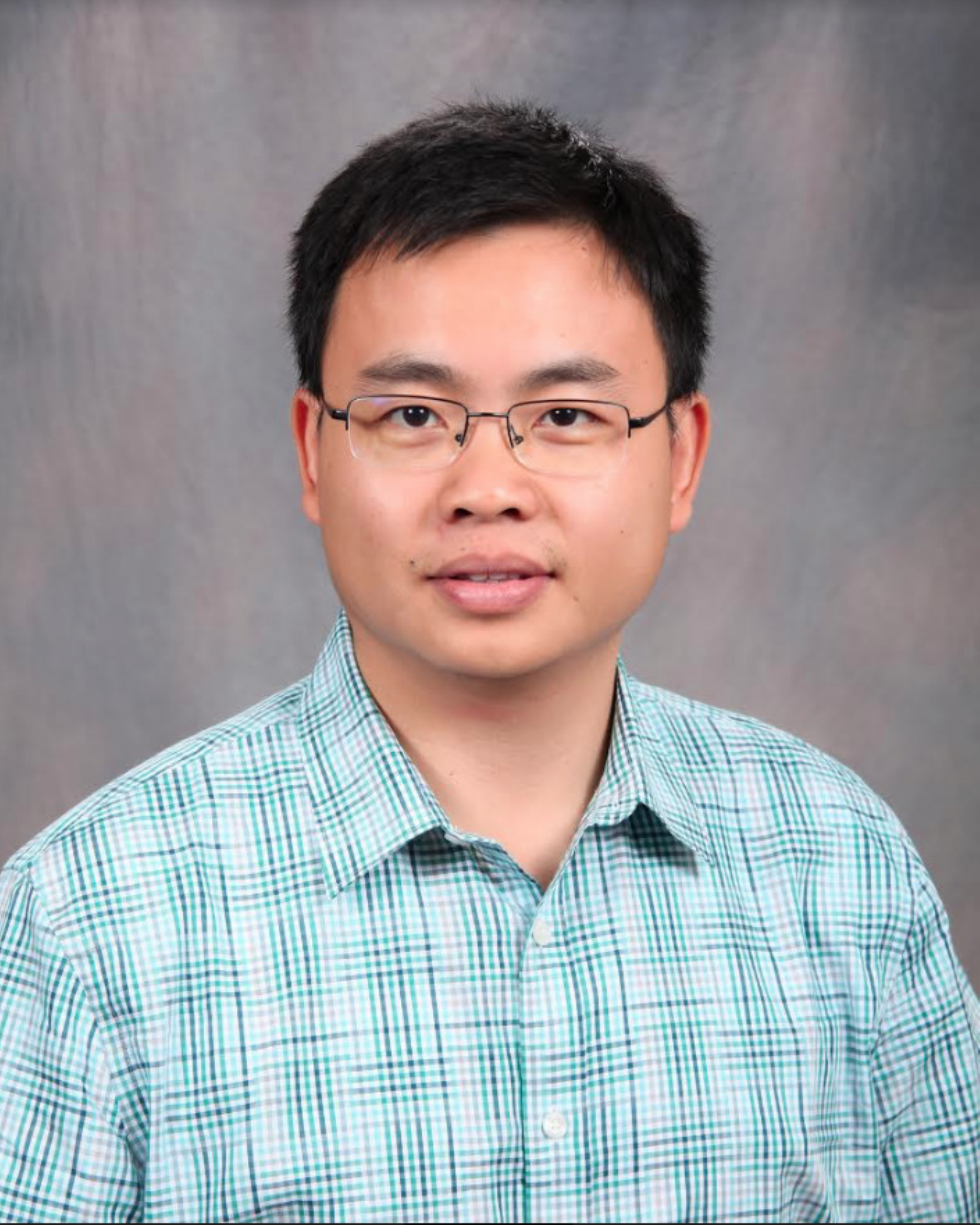}}]{Le Xie}
(S'05-M'10-SM'16) received the B.E. degree in electrical engineering from Tsinghua University, Beijing, China, in 2004, the M.S. degree in engineering sciences from Harvard University, Cambridge, MA, USA, in 2005, and the Ph.D. degree from the Department of Electrical and Computer Engineering, Carnegie Mellon University, Pittsburgh, PA, USA, in 2009.
He is currently an Associate Professor with the Department of Electrical and Computer Engineering, Texas A\&M University, College Station, TX, USA.
His research interests include modeling and control of large-scale complex systems, smart grids application with renewable energy resources, and electricity markets.
\end{IEEEbiography}

\end{document}